\documentclass[12pt,preprint]{aastex}

\begin{document}

\title{Morphological Evolution of Distant Galaxies from Adaptive
Optics Imaging}
\author{T.M. Glassman}
\email{glassman@astro.ucla.edu}
\author{J.E. Larkin \altaffilmark{1}}
\email{larkin@astro.ucla.edu}
\and
\author{D. Lafreni\`ere}
\email{dlafre@astro.ucla.edu}

\affil{Division of Astronomy, UCLA, 8371 Math-Sciences, Los
  Angeles, CA 90095} 
\altaffiltext{1}{Alfred P. Sloan Research Fellow}

\begin{abstract}

We report here on a sample of resolved, infrared images of galaxies at
z$\sim$0.5 taken with the 10-m Keck Telescope's Adaptive Optics (AO)
system. We regularly achieve a spatial resolution of 0.05$\arcsec$ and
are thus able to resolve both the disk and bulge components. We have
extracted morphological information for ten galaxies and compared their
properties to those of a local sample. The selection effects of both
samples were explicitly taken into account in order to derive the
unbiased result that disks at z$\sim$0.5 are $\sim$0.6 mag arcsec$^{-2}$
brighter than, and about the same size as, local disks. The
no-luminosity-evolution case is ruled out at 90\% confidence. We also
find, in a more qualitative analysis, that the bulges of these galaxies
have undergone a smaller amount of surface brightness evolution and have
also not changed significantly in size from z$\sim$0.5 to today. This is the
first time this type of morphological evolution has been measured in the
infrared and it points to the unique power of AO in exploring galaxy
evolution.

\end{abstract}


\section{Introduction}

Observational constraints on the growth of galaxies and the role of star
formation within them over time are finally within the reach of current
data. Powerful new techniques, thanks to modern space-based and
ground-based telescope facilities, have allowed researchers to make more
precise measurements of galaxy properties. In this paper we focus on one
aspect of this progress that is especially promising -- resolved images
of distant galaxies.

There has been a series of recent studies taking advantage of resolved
data to measure the evolution of disk properties, especially B-band disk
surface brightness to z$\sim$1 from Hubble Space Telescope (HST)
imaging.  These studies have provided some constraints on the amount of
evolution taking place, but there have been contradictory results,
mostly due to disagreements about the impact of selection effects.
\citet{schade} and \citet{roche} found B-band disk surface brightness
evolution of $\sim$1.6 mag to z=0.73 and $\sim$0.95 mag to z=0.9,
respectively, but neither focused on the impact of selection effects on
their results. \citet{lil} also found a high level of surface brightness
evolution ($\sim$0.8 mag to z=0.67), but calculated that as much as 0.3
mag of this could be due to selection effects.  \citet{vogt} obtained
rotation curves of galaxies to z$\sim$1 and found relatively mild
evolution in the Tully-Fischer relation ($\Delta M_B \lesssim$0.2 mag),
although their selection effects were very complicated.  \citet{simard}
concluded that almost all of the disk surface brightness evolution they
detected ($\sim$1.3 mag to z=1) is due to the selection effects. They
also reanalyzed the \citet{schade} and \citet{roche} studies and found
that selection effects could account for most of the evolution seen
there as well. Both \citet{vogt} and \citet{simard}, however, do detect
(even after accounting for selection effects) a population of galaxies
at z$>$0.5 that have higher surface brightness than the locus of nearby
galaxies.

The progress derived from this resolved data has, moreover, been limited
to observations of optical disk properties.  To date the only resolved
images of distant galaxies in the infrared (IR) have been those taken
with NICMOS on HST. However these data mostly provided global colors and
luminosities \citep[e.g.][]{tep}.  The resolution of the camera
($\theta=0.16\arcsec$) was insufficient to get reliable morphological
information and profile fitting routines failed more than 50\% of the
time \citep{corb}.  Observing in the IR reduces problems due to
morphological K corrections that can arise with optical images which
sample rest-frame ultraviolet (UV) light at high redshifts
\citep[e.g.][]{bunk}. IR data more directly sample the mass
distributions of distant galaxies since such images are less biased
towards regions of high star formation.

Although a picture of the structure and kinematics of nearby bulges is
beginning to emerge \citep{bur}, resolved bulge observations have also
been rare at high redshift. HST optical data does offer resolved images
of bulges to z$\sim$1, but most analysis of these data has been confined
to such things as bulge to disk ratios and bulge colors
\citep[e.g.][]{lil, ell}.

In order to study resolved structures, including bulges, we have started
a campaign to exploit high-resolution, IR images of distant galaxies
from the 10-m Keck Telescope's Adaptive Optics (AO) system.  These data
regularly have a resolution of $\sim$0.05$\arcsec$ in the H band, equal
to that of the HST optical data and three times higher than that of
NICMOS.  This resolution is sufficient to resolve bulges to arbitrary
redshifts.  The AO observations of the galaxies in our sample are
described in \S2 along with our methods of obtaining redshifts.  In \S3
we discuss the procedure used to extract galaxy morphological
information from the AO images and in \S4 we present the comparison of
these data to a local sample, taking into account the selection effects
for both samples. In \S5 we discuss the implications of our result for
galaxy evolution. Throughout this paper, we use H$_{0}$=65 km s$^{-1}$
Mpc$^{-1}$, $\Omega_M$=0.25, and $\Omega_{\Lambda}$=0.75.

\section{Observations}

\subsection{Sample Selection Using NIRC Imaging}
Our strategy for finding target galaxies appropriate for AO observations
involved searching near 26 bright stars using NIRC, a wide-field,
non-AO, infrared camera on the Keck telescope \citep{ms}. From these
fields, we selected a sample of $\sim$150 galaxies that were chosen to
fall within $30\arcsec$ of the bright stars and to have K$<$21.5 mag in
a $3\arcsec$ aperture. Many of the fields also contain faint, off-axis
stars that are useful for calibrating the point spread function (PSF) of
the AO observations. Objects found in these fields were named by the PPM
designation of their guide star plus the offset (in arcseconds) from
that star. Further details about these observations and the sample
selection are presented by \citet{lg}.

\subsection{Adaptive Optics Observations and Data Reduction}
From 1999 April to 2000 August, we observed 12 galaxies from this
IR-selected sample with the Keck AO system (see Table 1). The
observations through 1999 December were conducted with KCAM, the
first-light camera on the Keck AO system, which has a field of view of
$4.5\arcsec$ and a plate scale of $0.0175\arcsec$ per
pixel. Observations in 2000 June and 2000 August were made with SCAM,
the slit viewing camera on the NIRSPEC spectrograph \citep{mcl}, which
has a field of view of $4.4\arcsec$ and a plate scale of $0.0172\arcsec$
per pixel in AO mode. Except for one, the galaxies were all observed in
the H band (1.65 $\mu$m) with additional observations for many of the
galaxies in the J (1.27 $\mu$m), K (2.2 $\mu$m), and K$\arcmin$ (2.15
$\mu$m) bands.  The 2000 August observations were the only ones made
under non-photometric conditions.

The observations were dithered, moving the galaxy to each quadrant of
the chip, so that the median of the object frames could be used as a
sky.  More elaborate dither patterns could not fit in the small field of
view of the cameras used. A faint, off-axis star (with a guide-star
offset matched to that of the galaxy) was observed as close in time to
each galaxy observation as possible.  In Table 1 the PSF star used to
calibrate each galaxy observation is listed. In some cases, when no PSF
was a perfect match, the galaxy observation was calibrated using two PSF
stars that bracket the galaxy observing conditions (e.g.\ one is too
close to the guide star and one is too far away). The off-axis PSF stars
had full-width-half-maxima (FWHM) ranging from $0.05\arcsec$ to
$0.15\arcsec$ and strehl ratios of $1-20$\%.

The data were reduced with standard IR procedures including sky
subtraction, flat fielding, and dark subtraction.  Super skys were
generated from the science images by masking out the quadrant containing
the galaxy or PSF star and taking the median of a stack of unaligned
images. Bad pixels in the images were removed by replacing them with the
average of their neighbors. In some cases, there were differences in the
bias level between the four quadrants of the chip and/or the individual
rows. These variations were removed by adjusting the clipped mean
(calculated with the object masked out) of each quadrant/row to the
average value for the entire image.

In order to combine dithered images of the same object, the separate
frames had to be aligned. We found, through tests with the PSF images,
that header or input offsets were only accurate to within a few
pixels. Each object observed, including the galaxies, therefore had to
be bright enough to centroid on in each individual 5-10 minute
exposure. This is a strong limiting factor in the size and depth of
the current sample. The aligned images were finally combined by
averaging the frames together.

In the end, the 12 galaxies in the AO sample include all galaxies with
K$<$18.5 mag in a $3\arcsec$ aperture that were observable during the
times we were at the telescope. This is bright enough to ensure accurate
centroids and the sample is 100\% complete to this photometric limit in
the NIRC images. 

\subsection{LRIS Spectra and Redshift Determination}

In order to obtain redshifts for the galaxies in our sample, we took
spectra of eight of them with the Low Resolution Imaging Spectrometer
\citep[LRIS;][]{oke} on Keck in 1999 December, 2000 August, and 2001
July.  The other four objects were not accessible during these
times. Exposure times were 30-50 minutes and good continuum detections
were made of seven of the galaxies.

Following the approach of \citet{coh99}, the redshifts were
calculated manually from a visual determination of the wavelengths of
spectral features. This technique is justified by the very low
signal-to-noise (S/N) of the spectra, which would make it difficult for
more automatic techniques to work. Also, the construction of template
spectra for an automatic search would be difficult and not well
justified for such a small number of objects.

For each spectrum analyzed, the wavelengths of all the spectral features
were recorded, as well as the location of the 4000 \AA\ break, if
present. These features were then compared to the set of Hydrogen
Balmer, Ca II, [OII], and CH lines. A redshift was determined from the
best fit (lowest $\chi^2$) of the wavelength of the observed features to
the rest wavelength of the lines. When only a spectral break is present,
the redshift was obtained from the location of the break only. One
galaxy had a featureless spectrum, so no spectroscopic redshift was
obtained for it.

We are also able to put limits on the redshifts of the six galaxies
without spectroscopic redshifts because of the fact that the redshifts of
galaxies with K$<$18.5 mag are quite well constrained, as demonstrated
by infrared-selected redshifts surveys \citep[e.g.][]{coh96}. Due
to a conspiracy of low co-moving volume for low redshifts and the rapid
fading of galaxies at high redshifts, our whole sample has an expected
redshift distribution of z=0.6$\pm$0.3, with a small percentage of
galaxies expected at redshifts greater than z=1. This agrees quite well
with the average of our spectroscopic redshifts (z=0.55). We used this
information to assign a redshift probability distribution to the
galaxies without spectroscopic redshifts. This full probability
distribution was used in all of the analysis, but to simplify the plots,
the galaxies without spectroscopic redshifts are shown at z=0.6.

\section{Extraction of Morphological Components}

In order to compare our sample galaxies to local populations, we need to
characterize their basic morphological attributes -- i.e.\ separate them
into disks and bulges and find the properties of each.  The usual method
of extracting this information is to create model disks and bulges and
combine them to find the model that best matches the galaxy image. The
biggest difficulty in applying this method to our AO images is the low
S/N per pixel; this limitation prevents us from doing
a full 12-parameter, 2-dimensional analysis.  In order to reduce the
number of free parameters, in addition to increasing the S/N of the
data, we decided to fit the models to the data in one-dimensional radial
profiles. This approach has been used by many other authors and direct
comparisons were made between this and other techniques in a few cases
(e.g.\ \citealt{dj96}; hereafter DJ96).

We assumed that the bulges are round and that they can be characterized
by exponential profiles, which eliminates three parameters (the position
angle, inclination, and sersic index of the bulge). (Sersic profiles
were tested for the bulges, but we found that these did not fit as well
as exponentials.)  We then estimated the position angles and
inclinations of the disks by stretching them, while conserving flux,
until they appeared, by eye, to be round (eliminating two more
parameters). The central pixel of each galaxy was then determined by
centroiding, followed by inspection of the radial profile to verify that
it was smooth (eliminating a final pair of parameters).  This left just
five parameters to be determined by the formal $\chi^2$ minimization
routine: bulge peak surface brightness ($\mu_{oBulge}$), bulge scale
size ($h_{Bulge}$), disk peak surface brightness ($\mu_{oDisk}$), disk
scale size ($h_{Disk}$), and an additive sky background level. To
further increase the S/N, as well as the strehl ratios, of the images in
our sample, we focused on analyzing just the H-band images (except for
PPM91714+18-17 which was only observed in K$\arcmin$).

The errors on the galaxy profiles were derived by taking multiple
azimuthal averages of the images while varying the position angle,
inclination, and central pixel within reasonable ranges (typically
$\pm$5$\arcdeg$, $\pm$10\%, and $\pm$1 pixel respectively). Four more
profiles were created by taking the azimuthal average in four quadrants
using the best fit position angle, inclination, and center.  The
standard deviation of the mean of all of these profiles was then taken
as the error for each point. The final profiles were also binned into
6-7 radially averaged points (with the radial extent of the region we
averaged over approximately proportional to the radius) in order to
increase the S/N and to establish a smooth profile more representative
of what we actually know about the galaxy and less subject to the effect
of small fluctuations.

The small field of view, lack of true sky, and problems with the
detectors for the AO images caused some additional difficulties. One
issue was that the absolute level of the sky background in the AO images
was a significant source of uncertainty.  In order to put some
constraint on this value for each image, we calculated the flux of each
object in a $3\arcsec$ diameter aperture in both the pre-AO, NIRC image
and the AO image. We then adjusted the background level of the AO image
so that the magnitudes of the object in the two observations were the
same. If the NIRC data was taken under non-photometric conditions or
there was no NIRC data in the same band, the background level of the AO
image was estimated. However, this led to much higher errors for the AO
photometry.

Another issue is that the AO galaxy profiles only had reasonable S/N out
to $\sim1\arcsec-2\arcsec$ radius.  The result is that the
characteristics of the outer portions of the galaxies' disks are poorly
constrained by the AO images and the disk sizes are degenerate with the
background offsets. In order to better characterize the outer regions of
the disks, we fit models to both the AO images and the NIRC images of
the same galaxies.

\begin{figure}
\plotone{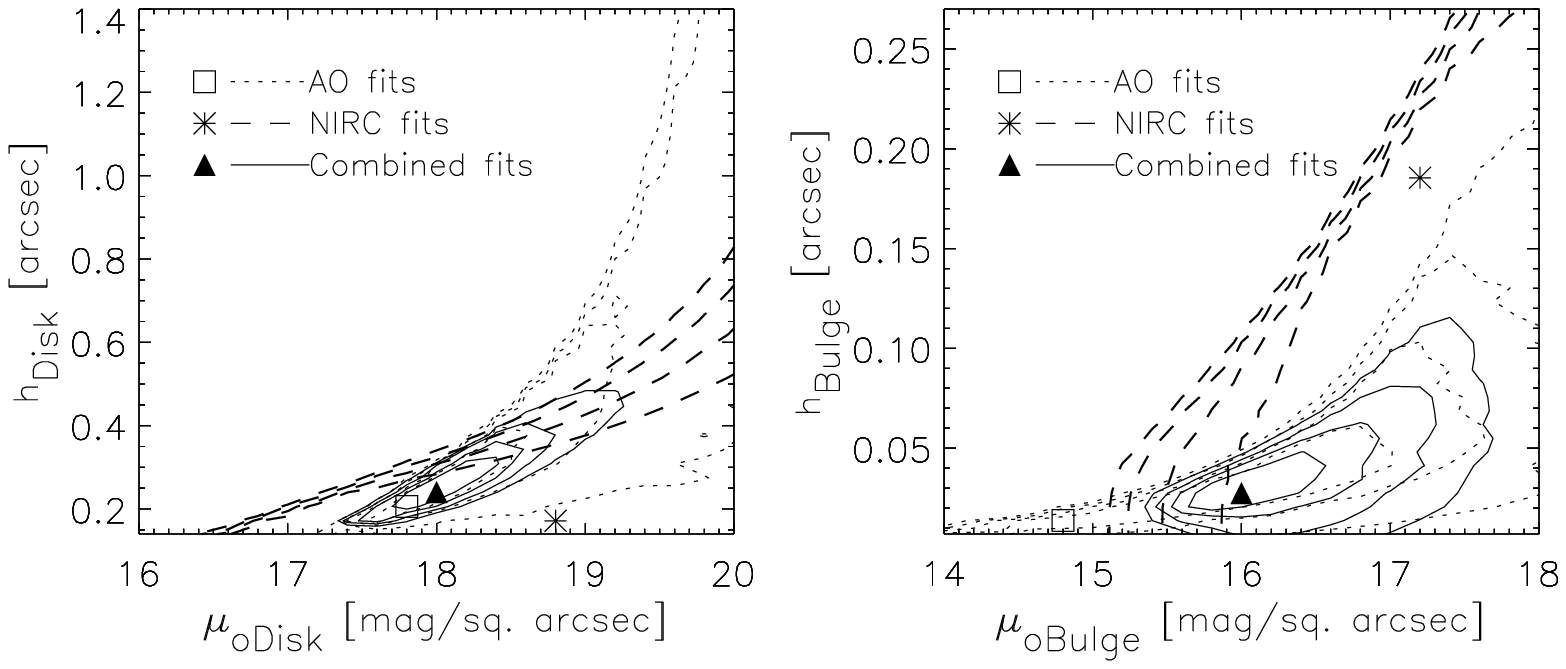}
\figcaption{A representative example of the results of the $\chi^2$
fitting routine (for the galaxy PPM101029+20-8) is shown. Contours of
$\chi^2$ at 1, 2, 3, and 4-$\sigma$ above the minimum are plotted with
respect to the four tested parameters (disk size and surface brightness
on the left, bulge size and surface brightness on the right). The three
sets of contours are for fits to just the galaxy profile from the AO
image, fits to just the galaxy profile from the NIRC image, and the sum
of the $\chi^2$ values for these two. The points mark the best fitting
model for each of the three cases.  As is fairly typical, the addition
of the NIRC information restricts the result to models with smaller disk
sizes than does the AO data alone, but it has less of an effect on the
bulge parameters.  This is consistent with the partial degeneracy
between the disk size and the sky background seen with the AO data, due
to the small field of view of these images.  Although each data set by
itself can lead to degeneracies in the model parameters, the two data
sets are complementary and therefore constrain the parameters quite
tightly when used together.
\label{chisq}}
\end{figure}

Model galaxies were created by making 2-dimensional, exponential disks
and bulges, convolving each one with the PSF, adding together a disk and
a bulge with a given flux ratio, and taking the azimuthal average of the
resulting image. The best fitting model for each data set was found by
minimizing $\chi^2$ between the disk/bulge combinations and the data.
For each set of model parameters, the $\chi^2$ values for the two data
sets were then added together to form a combined $\chi^2$.  As expected,
the NIRC images don't provide much of a constraint on the bulge
characteristics (especially their sizes) since the bulges are unresolved
in these images.  However, they do help constrain the disk
characteristics in ways that the AO images can't (see
Figure~\ref{chisq}).  The final model for each galaxy was taken as the
model with the lowest combined $\chi^2$ value. One-sigma errors were
found for each parameter by accepting models that produce a given range
of $\chi^2$ values ($\chi^2_{min}$+7.04 for six parameters).

\subsection{Models and Tests}

\begin{figure}
\plotone{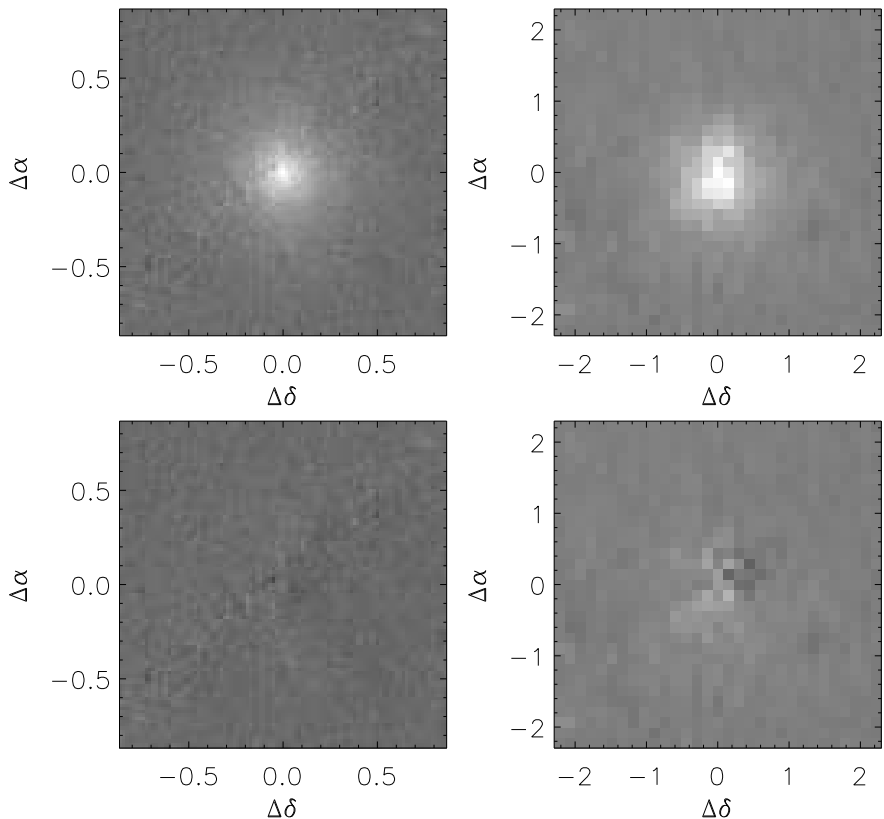}
\figcaption{Shown above are the AO (top left) and NIRC (top right) images
of the same galaxy (PPM102164+18-1), along with the residuals (bottom)
obtained when the best combined-fit model is subtracted from each. The
greyscales used to make these images are the same for the original and
the residual image in each pair.  It can be seen that there is
relatively little residual flux, although the NIRC residual is slightly
worse than the AO residual in this case: there is 20\% residual flux in
the NIRC image, as opposed to 7\% residual flux in the AO image. The
residuals in this case are representative of the rest of our sample.
\label{resid}}
\end{figure}

Various tests were conducted in order to confirm that our analysis
methods were functioning correctly.  We subtracted 2-D versions of the
best-fit model from each galaxy image in order to examine the residual
flux (see Figure~\ref{resid}).  For the five galaxies for which we have
tightly constrained fits, the residuals of the AO images contained less
than 10\% of the flux of the galaxy within a $3\arcsec$ aperture.  The
five with less well constrained fits still leave less than 30\% residual
flux in the AO images. These residuals are caused by a variety of
effects. For the brighter galaxies with better constrained models, the
primary cause is probably mismatches between the PSF in the galaxy image
and the PSF used in the analysis. Another cause of residual flux is
asymmetric structures or structures besides simple disks and bulges in
the galaxies themselves. For the fainter galaxies with less well
constrained models, large-scale noise fluctuations in the field are more
likely to provide the dominant contribution to the residual flux.
Several of the AO images of these galaxies have S/N as low as 3, which
means that the 30\% residuals could be simply due to noise in the
images.

A small Monte-Carlo test was conducted to assess the effectiveness of
the analysis routine in extracting the correct parameters.  We randomly
selected galaxy characteristics from a sample of local galaxies (DJ96)
and model galaxies were constructed using exponential disks and bulges
with these parameters and random position angles (PAs) and inclinations
for the disks.  A small random sky level offset was included and the
center of each galaxy was shifted randomly by a few pixels.  Each model
galaxy was then convolved with an AO and a NIRC PSF and residual images
from our data were added to each to create appropriate noise and sky
irregularities.  Each model galaxy was then analyzed in the same way as
our real galaxy images.  Because so much of the analysis routine is done
by hand, especially in finding the PA and inclination of the disk, only
four model galaxies were fully analyzed.  Still, this is enough to
estimate the accuracy of the method.  For the two models with small
inclinations (a/b$<$1.25), we found that the routine did a good
job. The derived parameters for these two objects, including the
inclinations, were all within 30\% of their true values and 3/4 of them
were within 15\%.  For the other two objects, which were closer to edge
on (a/b=1.3 and 2.3), the routine had less success.  We underestimated
the inclination for both of them (by $\sim$20\%) and all the other
derived parameters were therefore farther off from their input values
(3/4 of them were more than 30\% off).  However, the errors appear to
have a random nature; each parameter is underestimated and overestimated
with about the same frequency, indicating no systematic shift in the
derived parameters with respect to their real values.  To use what we
learned from these models about the accuracy of the analysis method, we
calculated extra error terms that, when added in quadrature to the
statistical 1-$\sigma$ errors for each parameter, accounted for the
difference between the input and calculated values of that
parameter. This extra error term was derived for each of the four model
galaxies and the mean of the four results was then added in quadrature
to the 1-$\sigma$ errors for the real galaxies.

\begin{figure}
\plotone{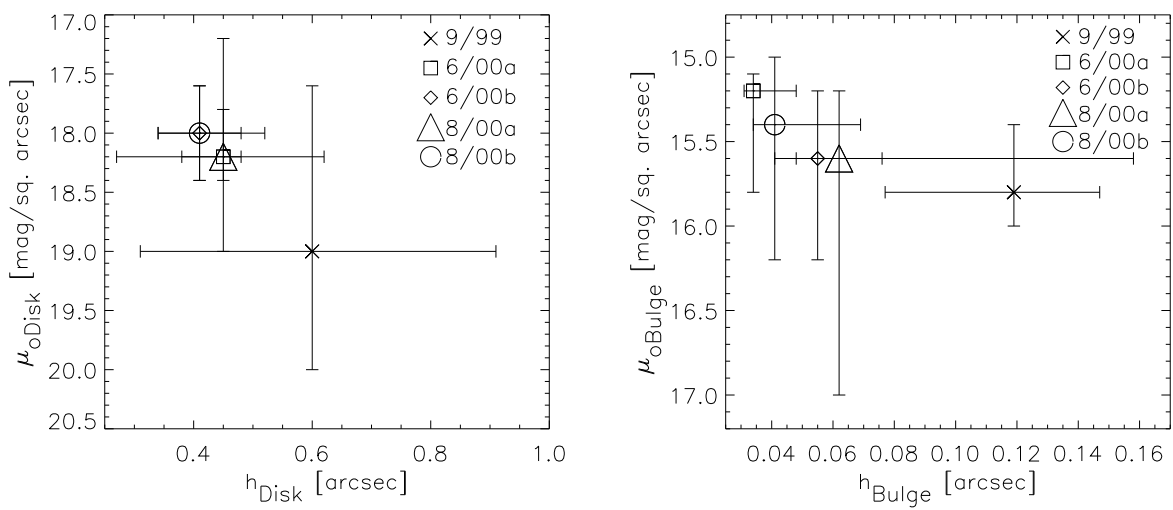}
\figcaption{Shown here is a comparison of morphological parameters derived
from five separate AO images of the galaxy PPM114182+6+27.  On the left
hand side are the disk parameters (size and surface brightness) and on
the right are the same parameters for the bulge.  The five disk models
are consistent within the 2-$\sigma$ error bars, as shown in the
plot. For the bulge, the only discrepant point is from 1999 September,
as expected.
\label{114182}}
\end{figure}

We also wanted to assess the effect of PSF variation on our results and
the repeatability of the derived morphological parameters. The galaxy
PPM114182+6+27 (which was imaged during three different runs, under a
variety of conditions) provided us with an opportunity to examine these
issues. The conditions in 1999 September were highly variable with the
strehl ratio of the off-axis PSF star changing from $\sim$9\% to
$\sim$2\% over a period of 45 minutes. We used only the best-quality PSF
and galaxy images in order to include the most information about the
galaxy structure and to use the best matching PSF possible.  This meant
throwing out four of the seven images of the galaxy from this night.
During the entire 2000 June run, the strehl was high ($\sim$10\%) and
stable and the conditions were photometric. We have two observations of
the galaxy from this run, taken on two different nights.  In 2000 August
the observations were not photometric and the seeing conditions, while
not offering a very high strehl, were stable.  There are again two
images from two different nights during this run.  We analyzed all five
of the images for this galaxy and found that the derived morphological
parameters were consistent within the 2-$\sigma$ error bars for all of
the observations (see Figure~\ref{114182}) and within the 1-$\sigma$
error bars for all but those in 1999 September. The June observations,
which were under the best conditions, had, as predicted, the smallest
error bars and the best $\chi^2$ values. As expected, the most
discrepant point is the bulge size derived from the September image,
when the conditions were most variable. This result gives us confidence
that the morphology derivations are robust and that we have correctly
modeled both systematic and random errors.

\subsection{Results and Details of Individual Objects}

\begin{figure}

\epsscale{0.71}
\plottwo{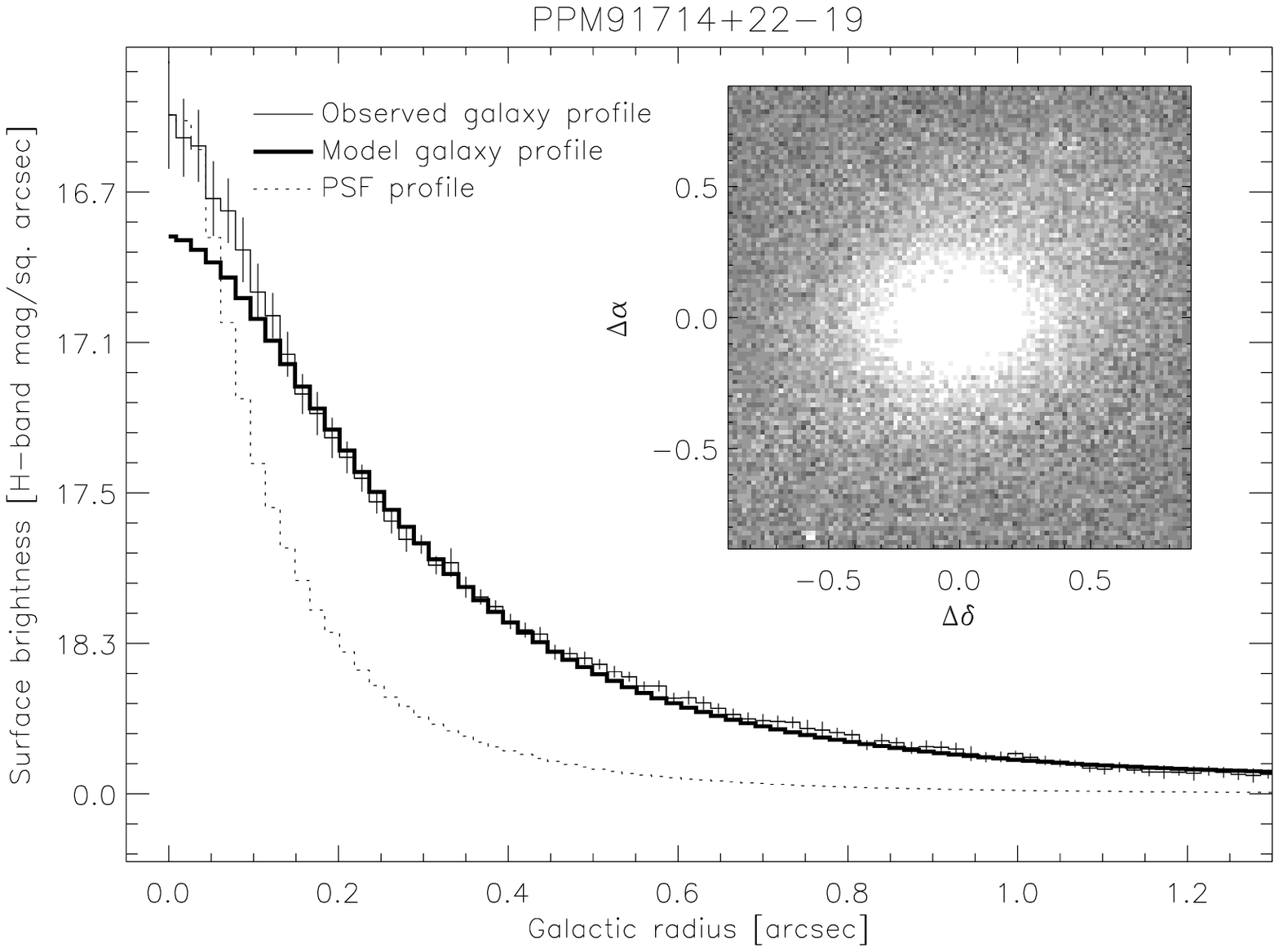}{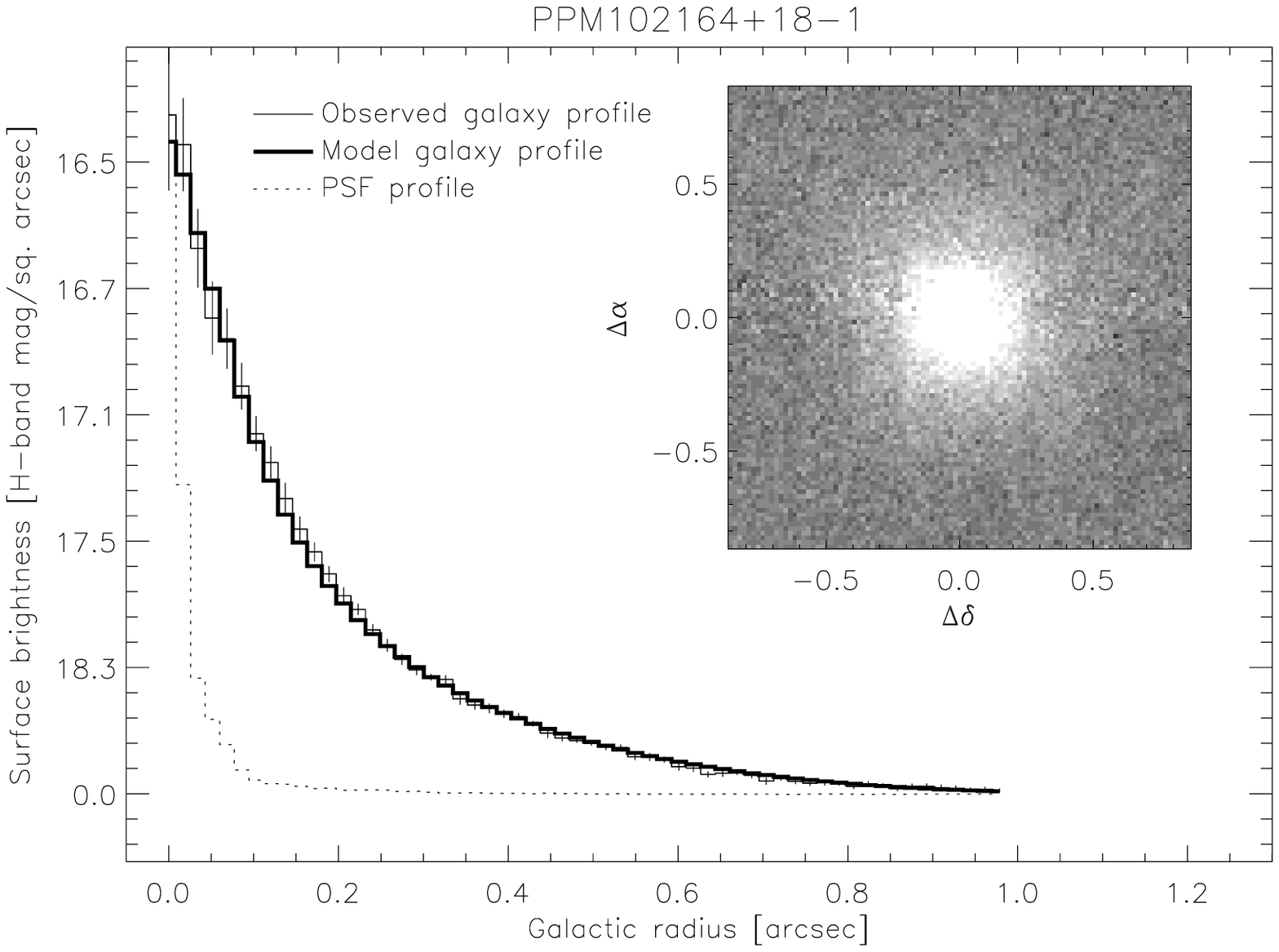}

\epsscale{1.6}
\plottwo{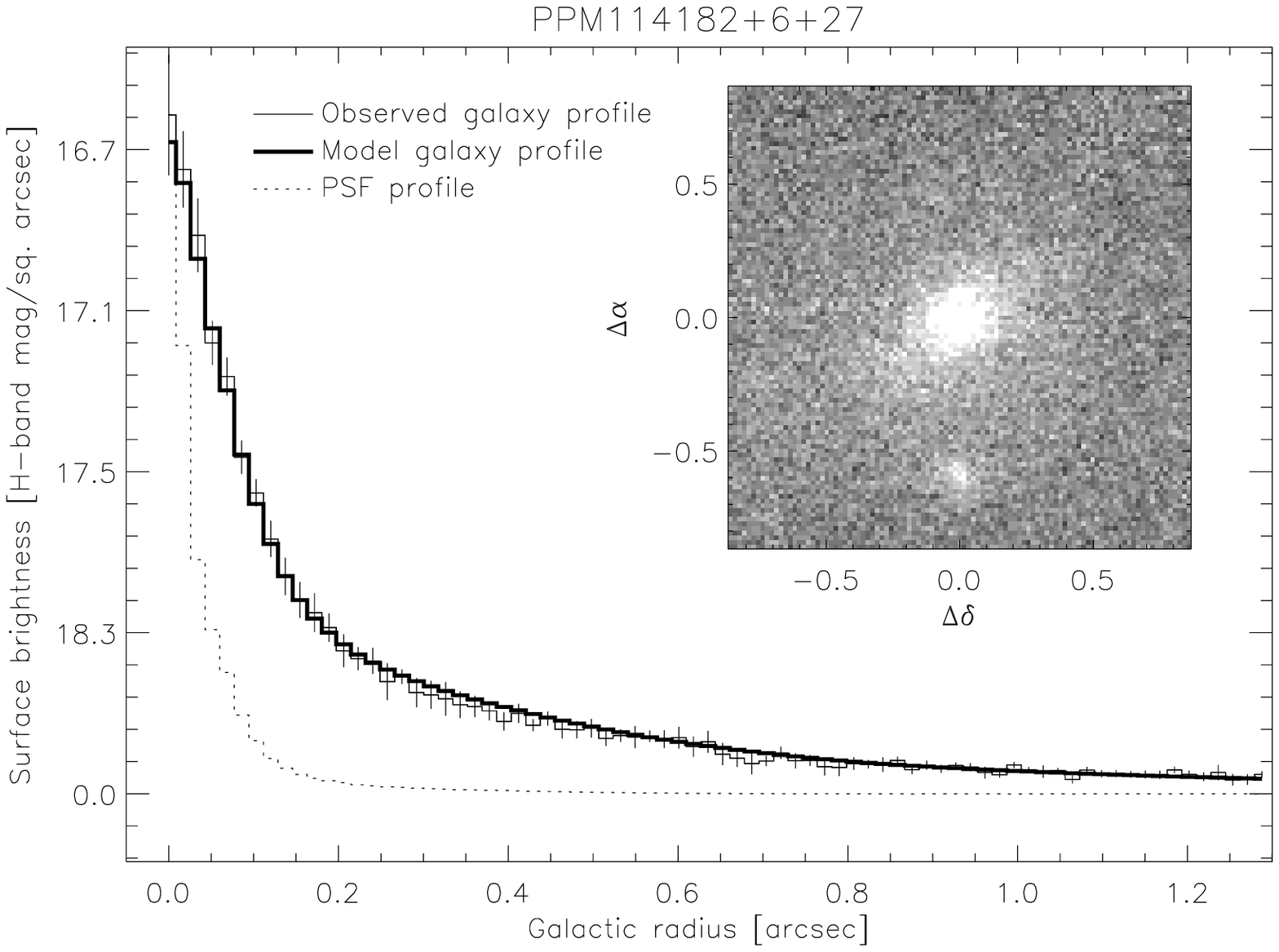}{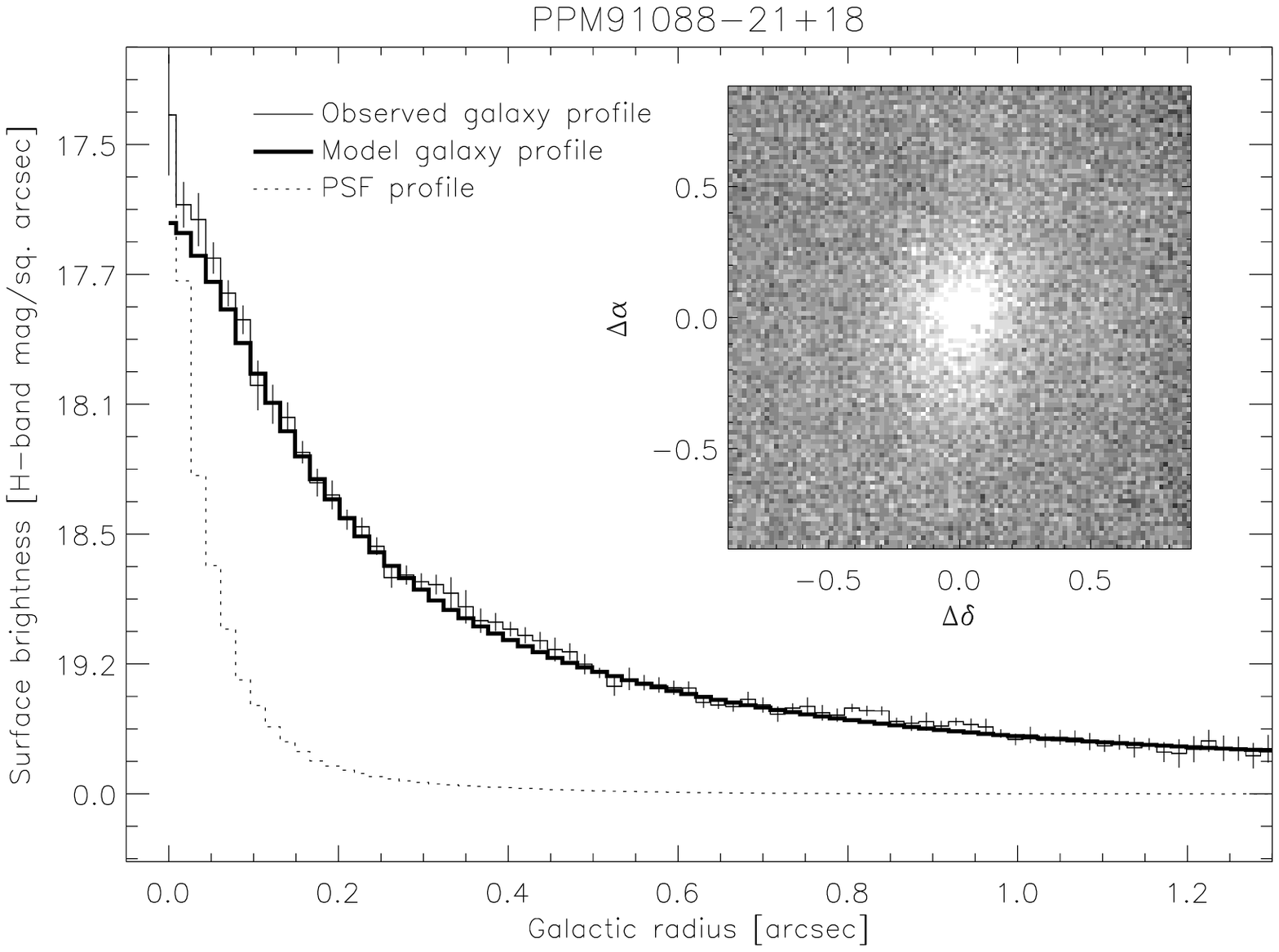}

\epsscale{3.49}
\plottwo{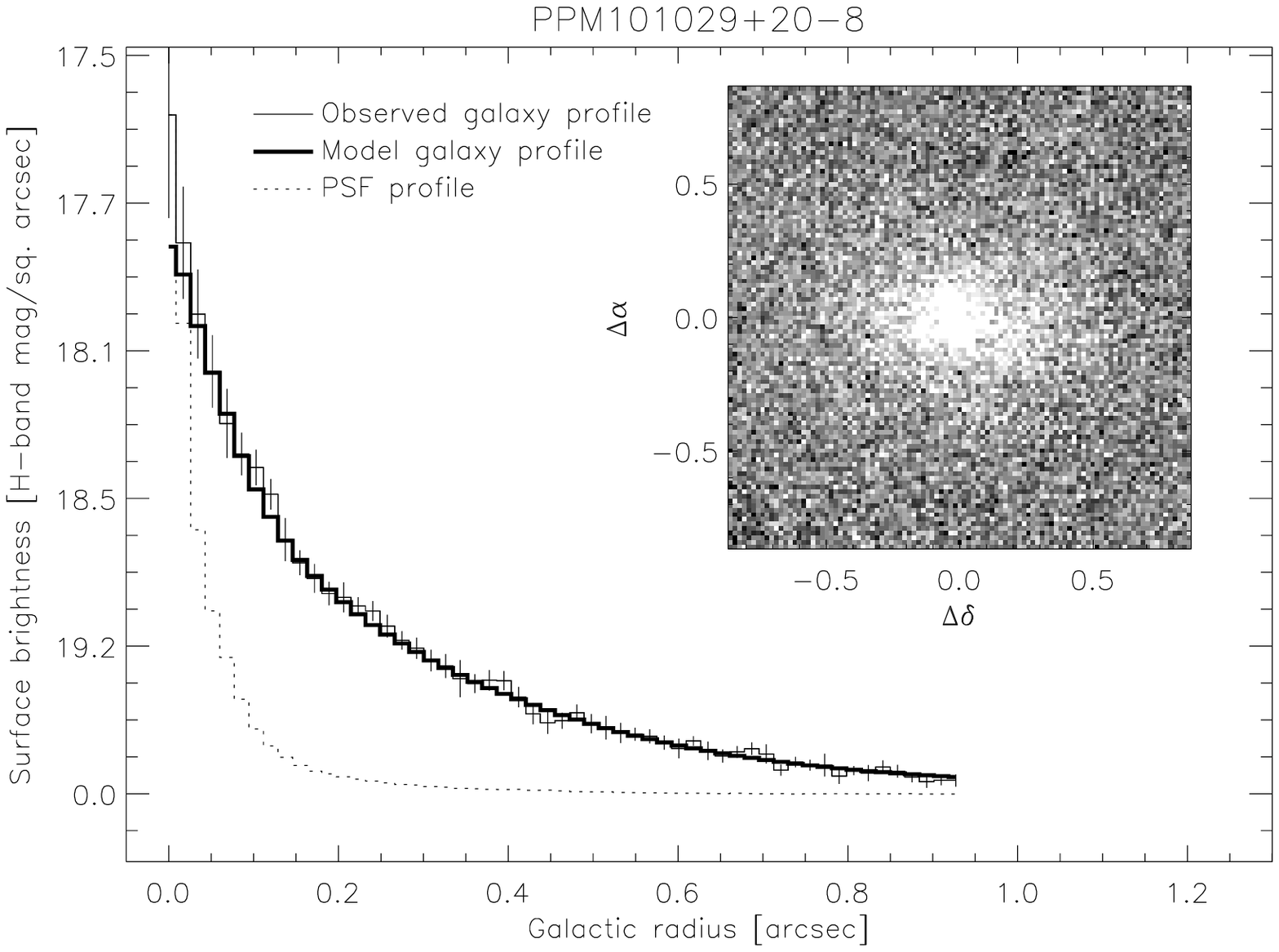}{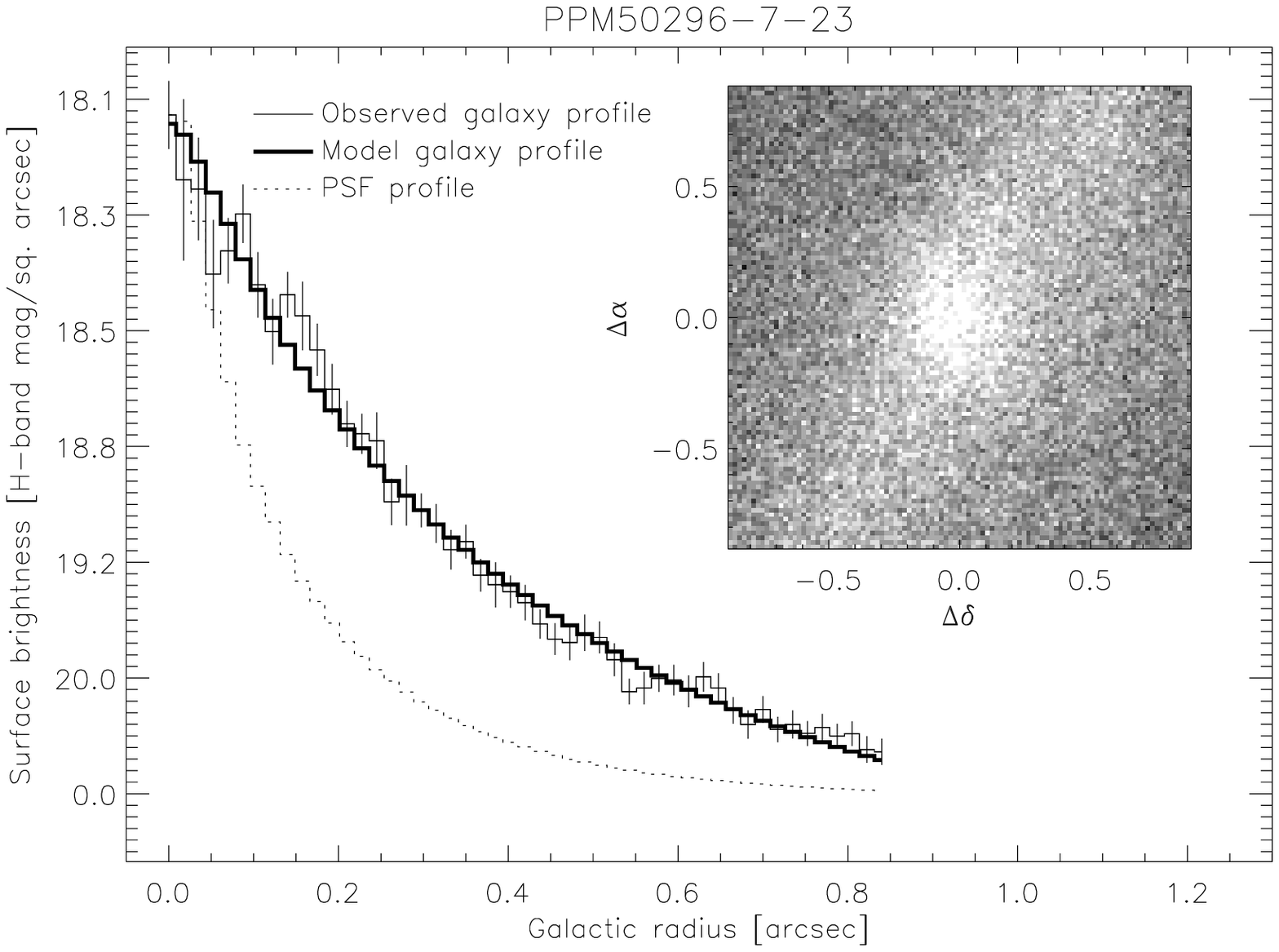}

\epsscale{7.9}
\plottwo{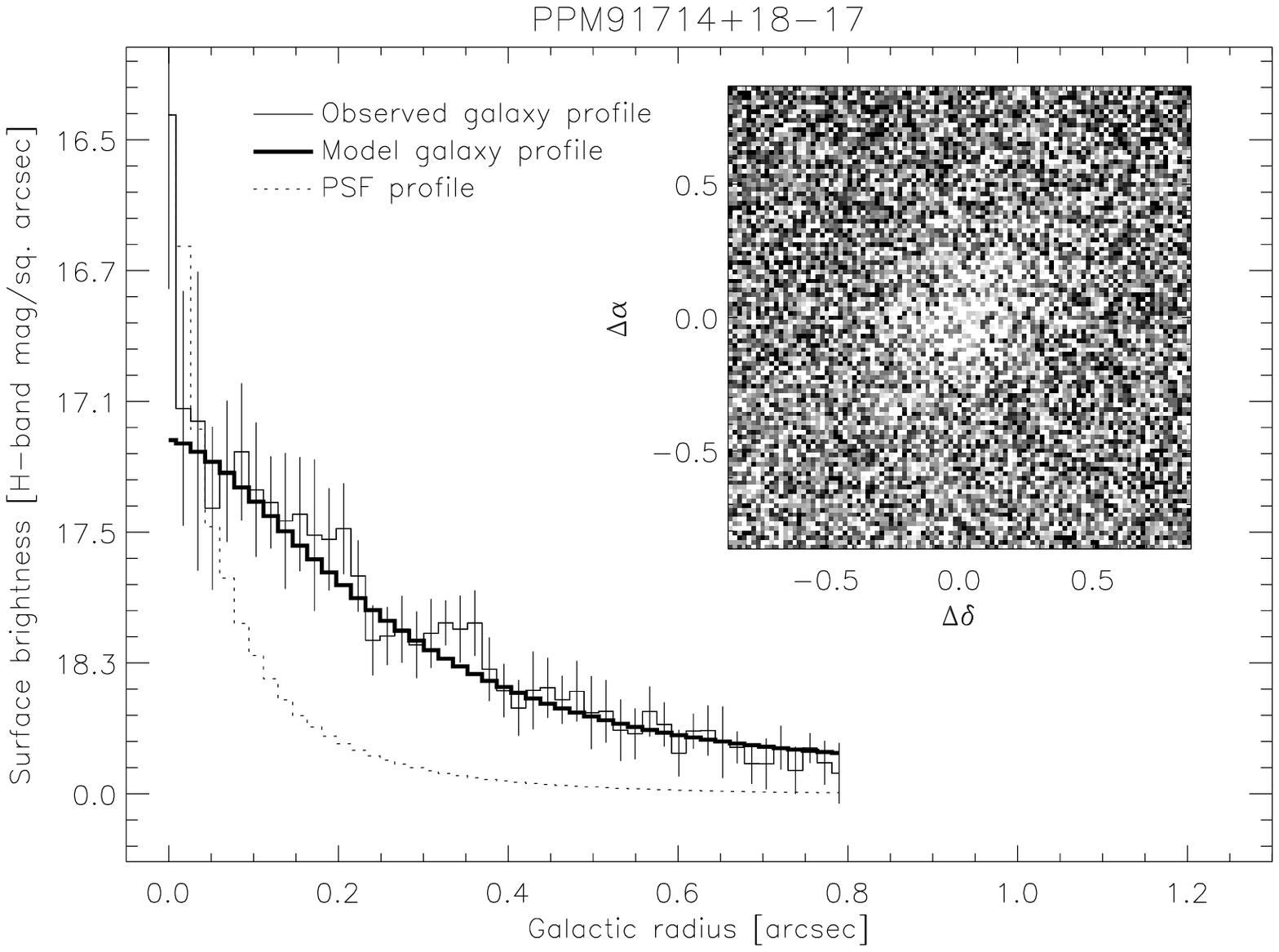}{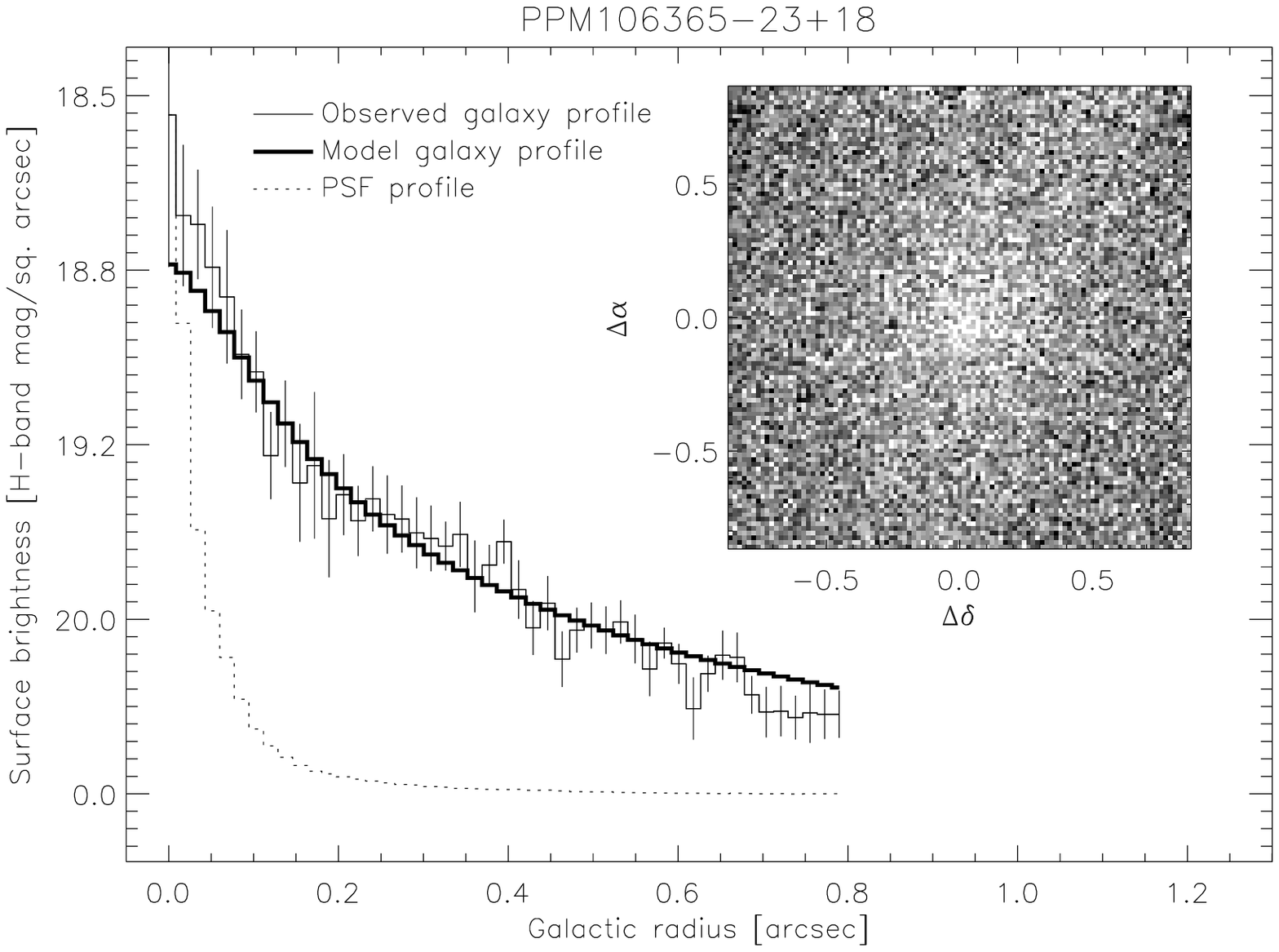}

\epsscale{17.0}
\plottwo{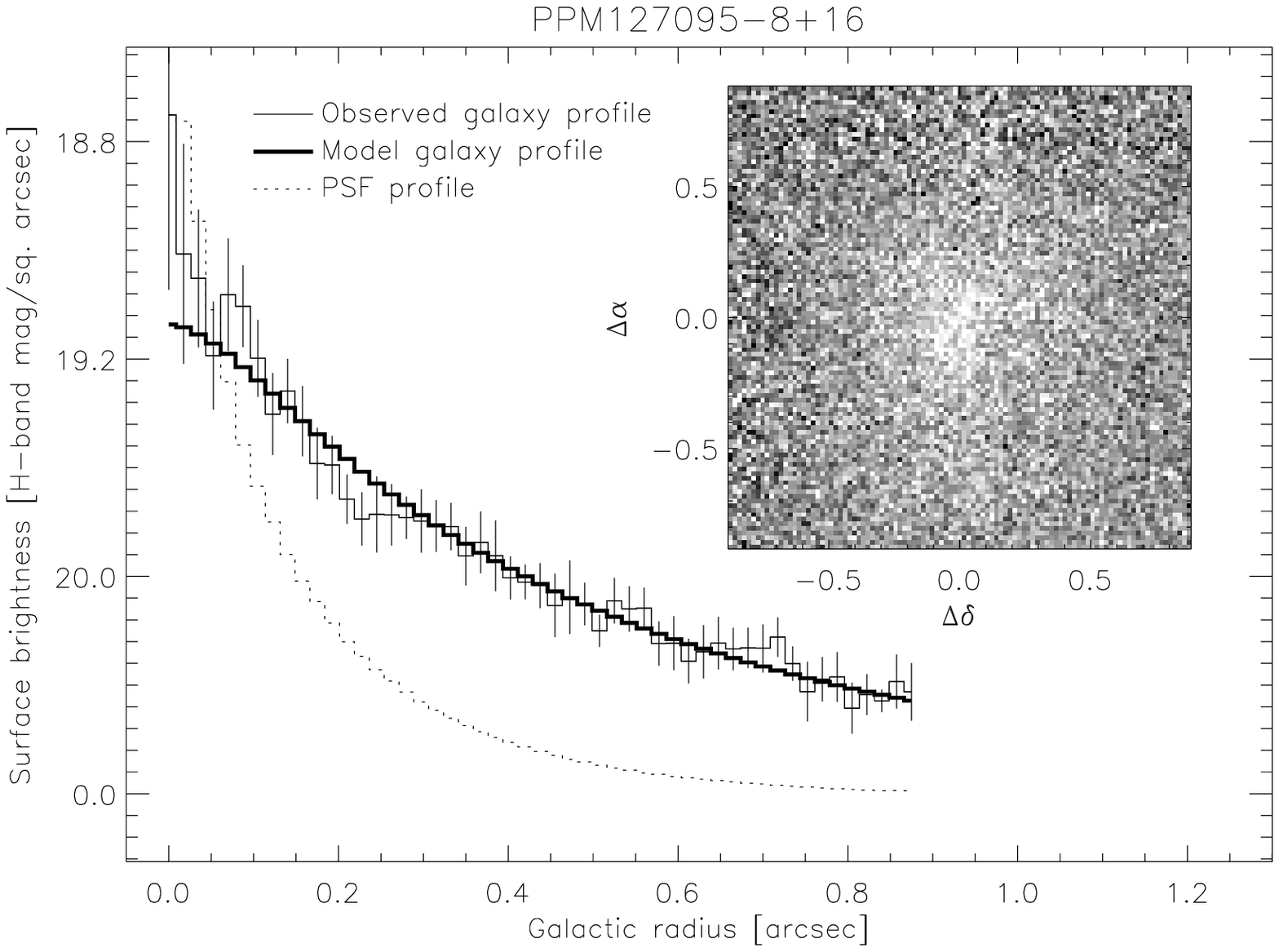}{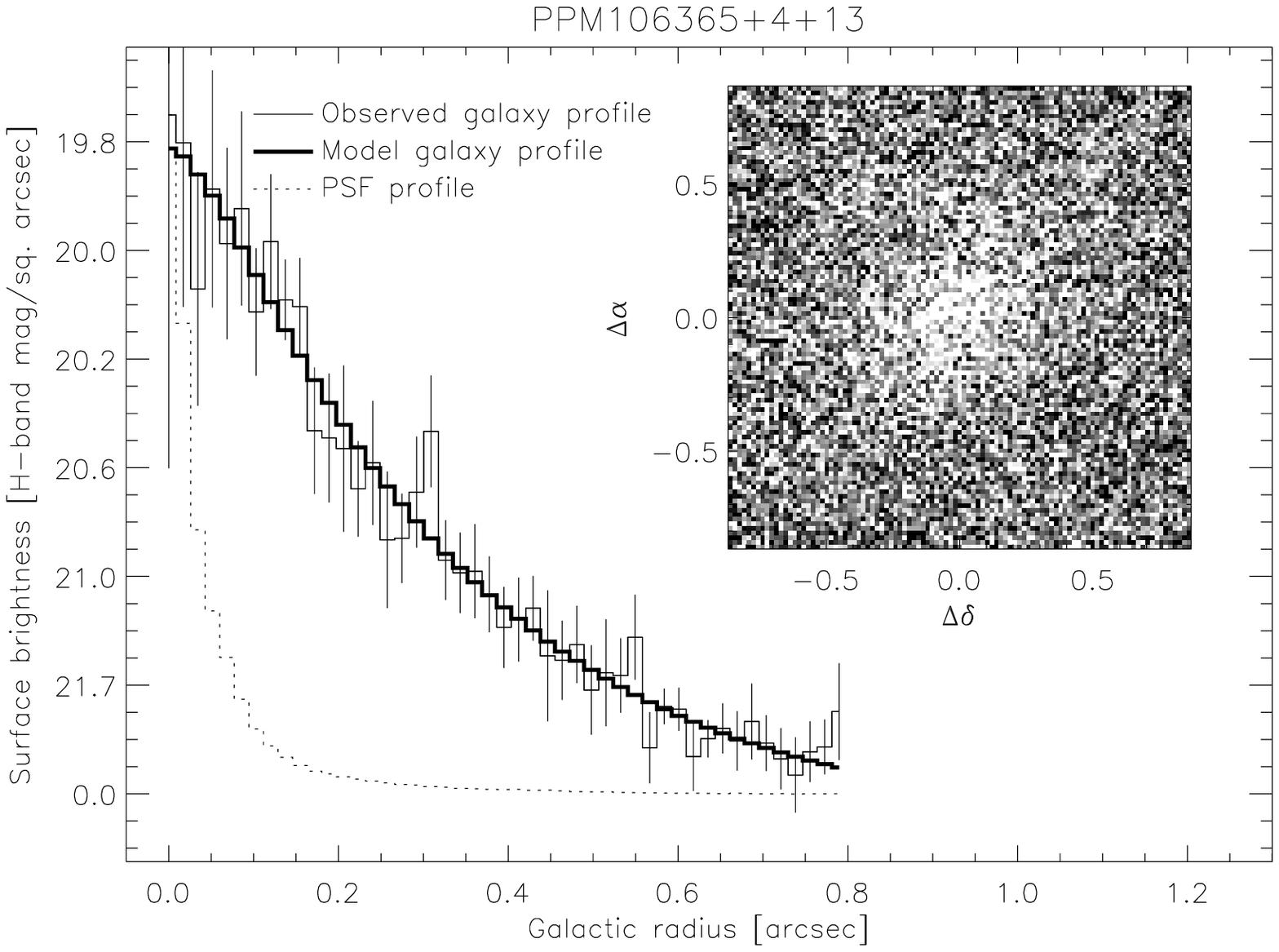}

\figcaption{The AO image of each galaxy that had a successful morphological
fit is shown here, along with profiles of the AO image of the galaxy,
the AO PSF, and the model that best fits the combination of AO and NIRC
data. The large variety in the data quality, the PSF shape, and the
model success can be seen here.  Note the off-axis point source to the
south of the galaxy PPM114182+6+27 and the strong bar in PPM50296-7-23.
\label{aofits}} 
\end{figure}

Models that fit the data well were found for ten of the 12 galaxies,
although five of the objects had large ranges of acceptable parameters.
For the five best constrained galaxy fits, the average uncertainties on
the four parameters were $\lesssim30\%$. For the other five, less well
constrained, fits the average uncertainties were $80\%-150\%$ (except
PPM50296-7-23 -- see below).  Although these uncertainties are large in
a fractional sense, they actually occupy a small fraction of the
available parameter space for normal galaxies. For two objects, no
acceptable fit was found (see below). Table 2 lists the best fit for
each successful case and Figure~\ref{aofits} shows the AO image of each
galaxy along with the galaxy, PSF, and model profiles.

\begin{description}

\item[PPM91714+22-19] This galaxy has a good fit to the one H-band image
and there is a small range of acceptable parameter space. The two
galaxies we observed in this field are $5.1\arcsec$ apart and both are
at z=0.46. It seems likely that they are interacting, although no
significant distortion can be seen in the current images.

\item[PPM102164+18-1] This galaxy has good fits to both H-band
images. There is a small range of acceptable parameter space for the
2000 June image and a larger range for the 1999 December observation.
The final model was obtained from a $\chi^2$ weighted average of these
two results. The galaxy is at a redshift of z=0.55.

\item[PPM114182+6+27] The best images of this galaxy were from two
separate observations in 2000 June (see above). The $\chi^2$ weighted
average of the results from these two observations was therefore used
as the final model, providing a good fit with well-constrained
parameters.  The galaxy is at a redshift of z=0.48, but this was based
only on the 4000\AA\ break, so it is less certain than the other
redshifts ($\Delta z=\pm0.05$).

This galaxy is unique in our sample in that the AO image shows a point
source $0.6\arcsec$ south of the nucleus, with an apparent H-band
magnitude of 22 (see Figure~\ref{aofits}). The knot then has an absolute
H-band magnitude of -20 and may be a very luminous star-forming
region. Similar ``super star clusters'' are seen in local galaxies, but
this would be among the most extreme examples and is about 2 mag
brighter than the brightest knot in the interacting galaxies NGC 6090
\citep{din}. It is likely that the knot in PPM114182+6+27 is
composed of many such regions or that it is a second galactic nucleus,
indicating an ongoing merger. It could also simply be a companion galaxy
seen in projection against the disk, although the projected separation
is very small for this to be the case. 

The probability of the knot being a faint Galactic star seen in
projection against the disk is quite small. There is $\sim$1\%
probability of a star of comparable magnitude being found within a
1$\arcsec$ radius of this galaxy \citep{mcoh}. This means that the
chances of finding a star aligned with any galaxy in our sample is
$\sim$12\%.

\item[PPM91088-21+18] There is a good, well-constrained fit to one of the
H-band images of this galaxy (from 1999 September), but the fit to the
other image (from 1999 December) is of lower quality (higher $\chi^2$
and bigger errors) due to a much shorter exposure time. Therefore only
the 1999 September result was used. This galaxy is at a redshift of
z=0.70.

\item[PPM101029+20-8] There is a good fit to the one H-band image of
this galaxy, with a small range of acceptable parameter space. It was
not observed with LRIS.

\item[PPM50296-7-23] This is the only galaxy in our sample with a strong
bar and we therefore could not analyze in the usual way.  Instead, we
took the azimuthal average in two 120$\arcdeg$ sections perpendicular to
the bar.  The disk was assumed to be face-on (i.e.\ no further
correction for this was added) and the rest of the analysis proceeded in
the normal way.  The analysis resulted in a good fit with reasonable
error bars on the parameters (average of 31\%), however some of the
light from the bar is certain to be contaminating the fit.  For this
reason the results are not considered to be of the best quality. This
galaxy was not observed with LRIS.

\item[PPM91714+18-17] There is a good fit to the only image of this
galaxy, which is in the K$\arcmin$ band. There is, however, a large range
of acceptable parameter space -- both disk parameters are only upper
limits. The two galaxies we observed in this field are $5.1\arcsec$
apart and both are at z=0.46. It seems likely that they are interacting,
although no significant distortion can be seen in the current images.

\item[PPM106365-23+18] This galaxy has a good fit to the one H-band
image, although there is a fairly large range of acceptable parameter
space.  The galaxy is at a redshift of z=0.60.

\item[PPM127095-8+16] This galaxy has a good fit to the one H-band
image, but there are rather poor constraints on the parameters and the
AO and NIRC models disagree by a larger than average amount
($\gtrsim$60\%). The galaxy was not observed with LRIS.

\item[PPM106365+4+13] There is a good fit to the one H-band image of
this galaxy. However, there is a very large range of acceptable
parameter space. Both disk parameters and the bulge peak surface
brightness have only upper limits and the bulge size is not constrained
at all. It is the faintest galaxy in the sample at K=18.4 mag.
Continuum light was detected in the LRIS spectrum of this source, but no
features were seen that could be used to derive a redshift.

\item[PPM102164+9+23] Because of the very short exposure time, the one
AO image of this galaxy only had enough S/N to create a profile out to
$\sim0.5\arcsec$ radius, instead of the $1-2\arcsec$ radius typical for
the AO images.  Because of this, the AO profile could not be used to
establish the photometry in place of the non-photometric NIRC image, nor
was it useful in constraining the morphology of the galaxy. This galaxy
was not included in our sample for the following evolutionary
analysis. Since its exclusion is primarily based on the short exposure
time, we do not consider this a significant bias to the sample. This
galaxy was not detected in an LRIS spectrum.

\item[PPM115546-4+6] This galaxy proved impossible to analyze
successfully because a diffraction spike from the guide star passed very
close to the galaxy in every image.  This makes it impossible to get an
uncontaminated profile of the galaxy, especially since the relative
placement of galaxy and spike shifts with each image. At a separation of
only $7\arcsec$, this galaxy is the closest to its guide star in the
sample. It was not included in the following evolutionary
analysis, but this should not bias the results. This galaxy was not
observed with LRIS.

\end{description}

\section{Comparing Local and Distant Galaxy Properties}

The ten galaxies for which we have derived morphological parameters (all
but PPM102164+9+23 and  PPM115546-4+6) represent a flux limited sample
with K$<$18.5 mag within a $3\arcsec$ aperture. We next want to compare
these properties to the same values for local galaxies.  We accomplish
this by creating a local comparison sample based on DJ96's observations
of 86 local disk galaxies. The morphological parameters of the galaxies
in this sample were measured across a wide range of wavelengths using
several different techniques -- including fitting exponential disks and
bulges to 1-D profiles in the H and K bands.  DJ96 also provided an
analysis of the selection effects present in their sample and an
estimate of the volume-limited, bivariate, K-band distribution function
of local disk properties. The main difficulty in using the DJ96 sample
is that it contains primarily Sb- to Sd-type galaxies. Our galaxies are
not inconsistent with this range of Hubble types, but we cannot properly
determine type and there may be early-type disk galaxies in our
sample. As discussed in \S5, we don't believe this has a significant
effect on IR surface brightness.

\begin{figure}
\plotone{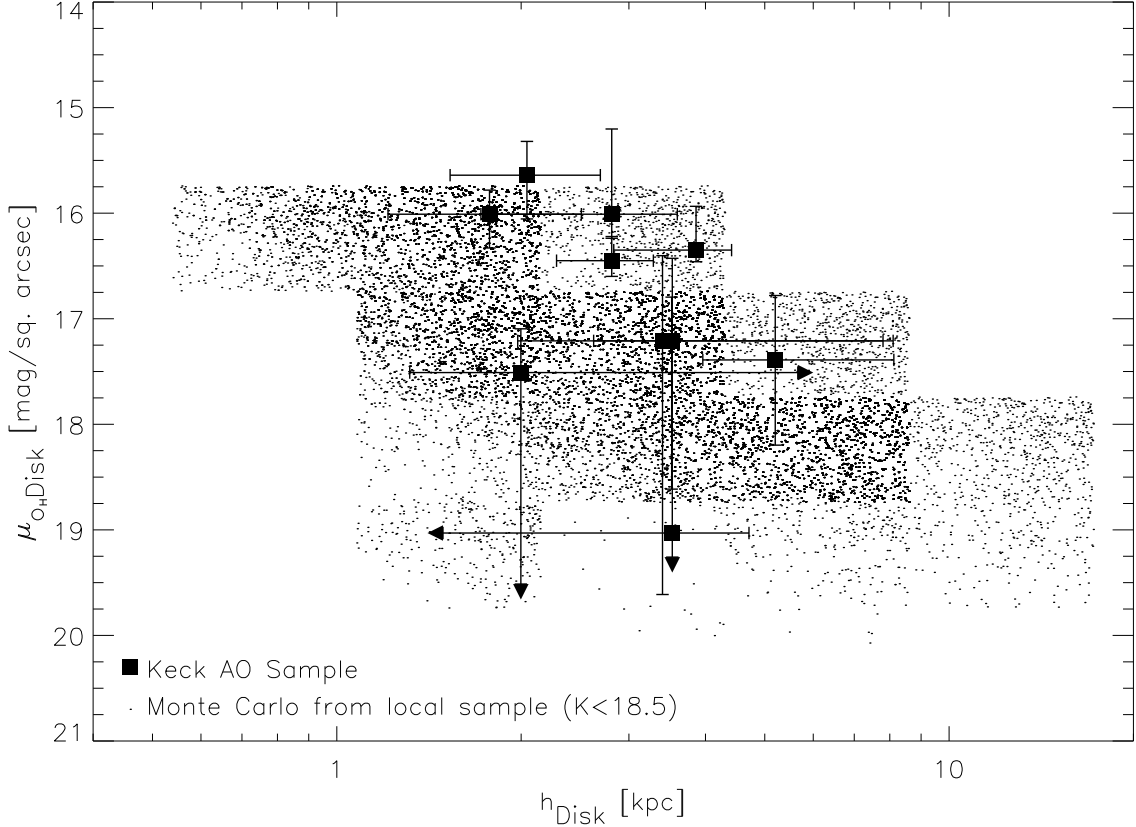}
\figcaption{Plotted here (small dots) are the H-band disk properties
expected for our galaxy sample if there is no evolution in the
population, based on the properties of a local galaxy survey (DJ96) and
the selection effects of both samples.  The boxes with error bars show
the same properties for the galaxies in our sample. The galaxies in our
sample without known redshifts are shown here as if they were at z=0.6
but the full allowed range of redshifts was taken into account in the
quantitative analysis. The edges in the local sample distribution come
from the discrete method used by DJ96 to model their volume correction.
\label{ddplot}}
\end{figure}

Starting with the volume-limited distribution function provided by DJ96,
we used Monte-Carlo techniques to create a large, random sample of
galaxy disks. We then found the highest redshift for each galaxy that
would allow it to be included in our data set (based on our selection
criterion of K$<$18.5 mag within a $3\arcsec$ aperture and taking into
account the K-band K correction versus redshift; \citealt{man})
and calculated a probability for each galaxy that is proportional to the
comoving volume out to this limiting redshift. This probability was used
to create a smaller, selected sample which represents our data set if
there is no evolution present. The rest-frame properties of these
objects were then adjusted from K-band to H-band surface brightness
using H-K=0.24 \citep{man} in order to directly compare them
to our H-band AO data. Figure~\ref{ddplot} shows the disk properties of
this Monte-Carlo, no-evolution, H-band sample compared to these same
properties for our sample.

The figure suggests that there is evolution between the two samples, in
the sense that a larger fraction of our population than of the
Monte-Carlo sample is at high surface brightness ($<$16.5 mag
arcsec$^{-1}$).  To formally compare the two and quantify this
evolution, we used the 2-D Komolgorov-Smirnov test \citep[K-S;][]{press}
which measures the probability of two samples being drawn from the same
parent population.  We used Monte-Carlo techniques, taking into account
the error bars on our data, to derive a probability distribution
function for the K-S value that is specific to our data set. Based on
this distribution function and the K-S value between our measured data
and the no-evolution sample, we found that there is $<$10\% chance of
getting our data if there is no evolution in the galaxy population.

\begin{figure}
\plotone{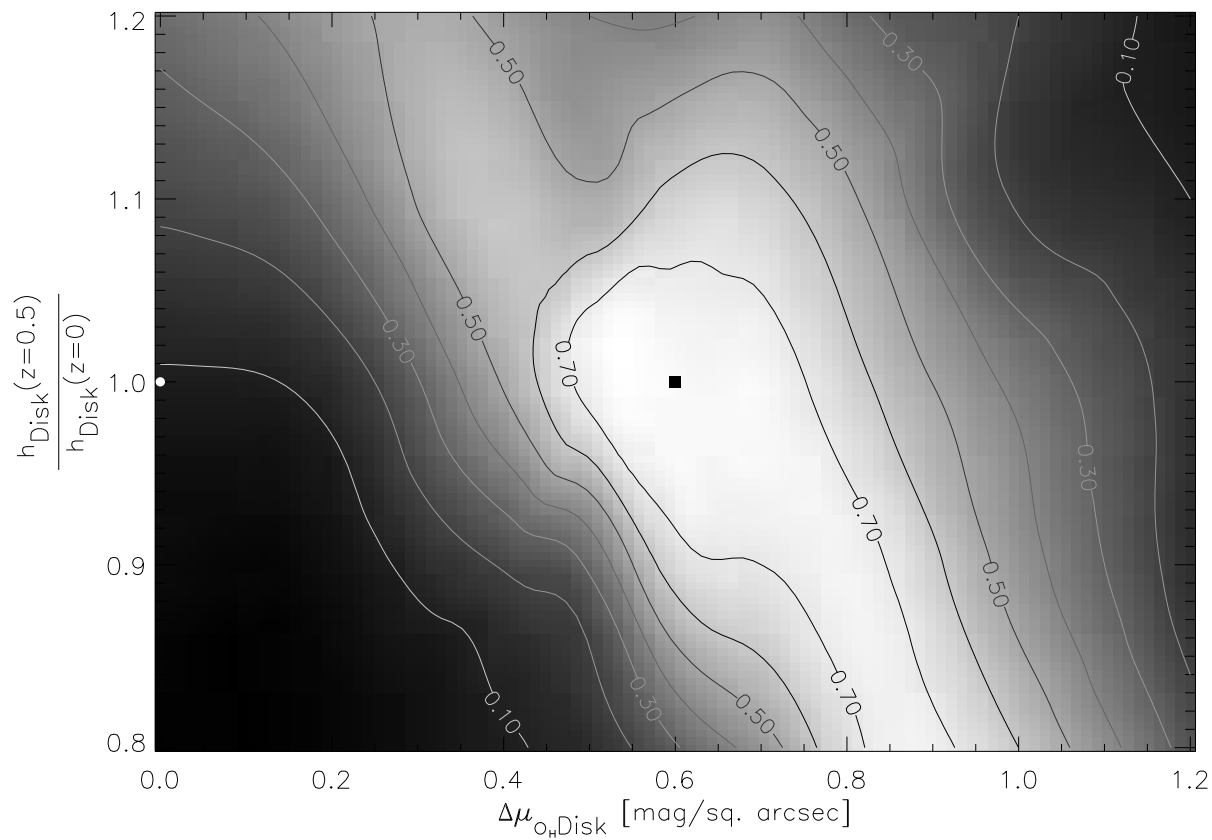}
\figcaption{Contours and greyscale of the probability of getting our AO
data given the varying amounts of disk evolution are shown.  The axes
are magnitudes arcsec$^{-2}$ of surface brightness evolution and
fraction of size evolution.  The best agreement between the two samples
occurs if the disks in the distant sample are 0.6 mag arcsec$^{-2}$
brighter than and about the same size as local disks (marked in the plot
by a black square).  Note that the no-evolution case is at the
middle-left side of the plot (white circle) and is ruled out at the 90\%
confidence level.
\label{evol}}
\end{figure}

To test what type of evolution is most consistent with the two measured
populations, we shifted the size and peak surface brightness of all the
galaxies in the Monte-Carlo sample created from DJ96's distribution
function. We then reapplied our selection effects to create several
random samples with varying amounts of evolution and used the K-S test
to compare each one to our data.  This is a very simple model of galaxy
evolution in that all galaxies change by the same amount, however it
does provide a way to measure the shift in the average properties
between the two populations. As Figure~\ref{evol} shows, the best match
occurs if galaxy disks at moderate redshifts were $\sim$0.6 mag
arcsec$^{-2}$ brighter than and about the same size as local disks. This
amount of disk evolution is consistent with our data at the 75\%
level. The no-evolution case is in the middle-left edge of the plot and
is strongly ruled out. There is some degeneracy seen in this plot due to
the weak correlation of size with surface brightness in the galaxy
population. The result is that we are more sensitive to evolution that
changes the average luminosity of the galaxies than to evolution along
lines of constant luminosity. If we accept all models that match our
data with at least a 65\% probability, we conclude that the disks of
galaxies to z$\sim$0.5 have evolved by $\sim0.6^{+0.3}_{-0.1}$ mag
arcsec$^{-2}$ in surface brightness and by a factor of 0.8 to 1.05 in
size.

\begin{figure}
\epsscale{1.1}
\plottwo{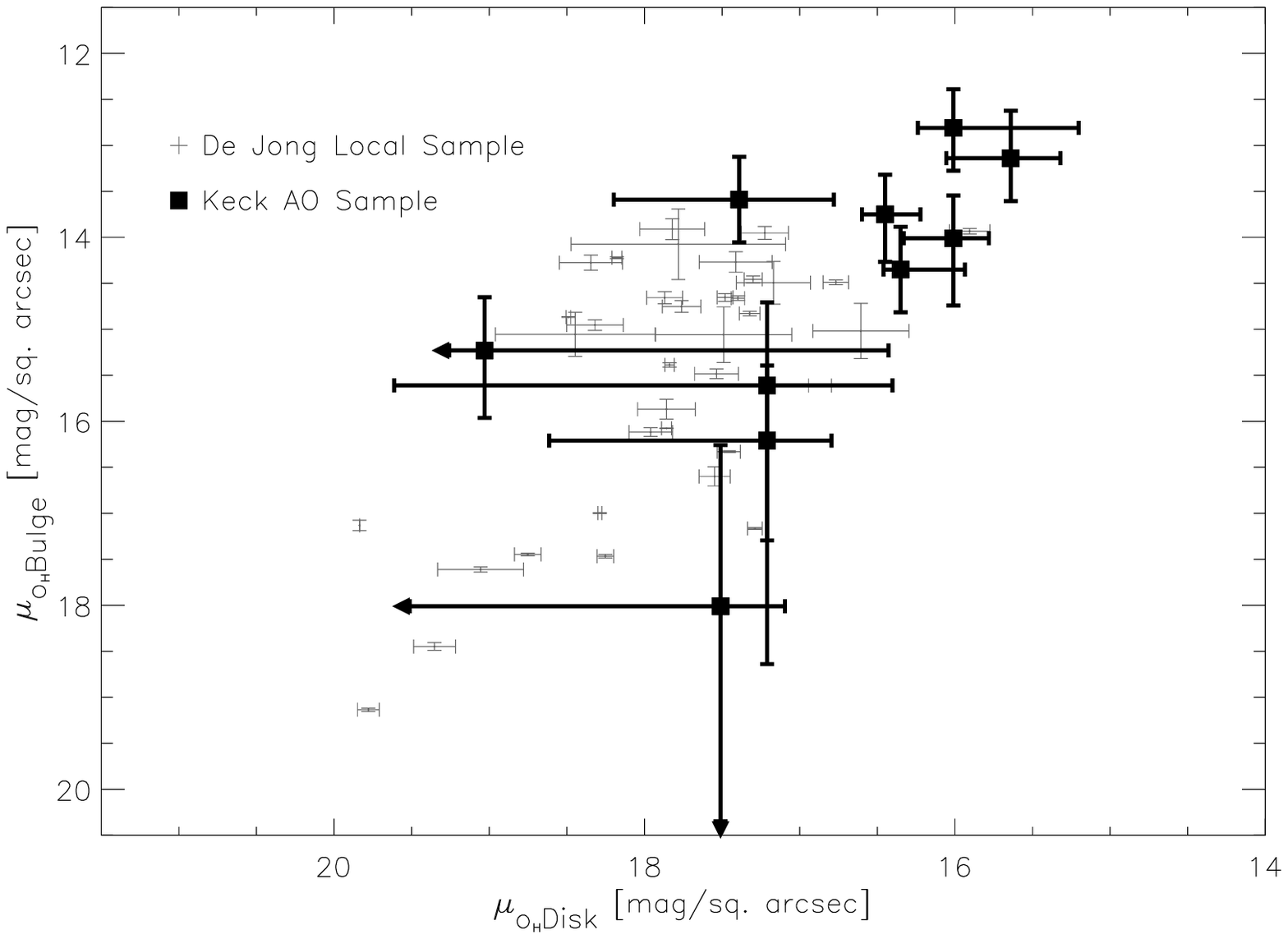}{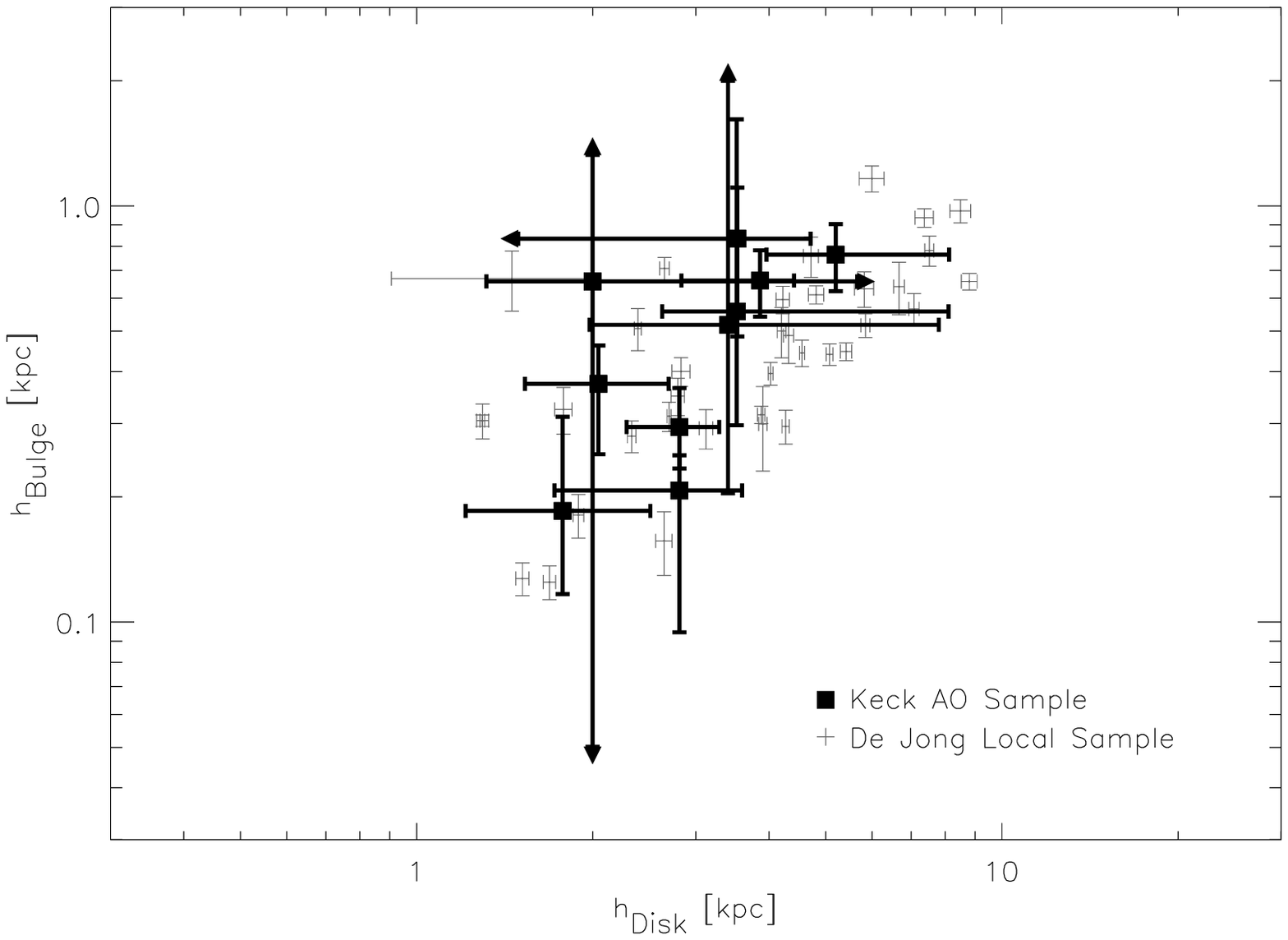}
\figcaption{The surface brightness (left) and size (right) of the disks and
bulges of our galaxies (bold) and a select sub-sample of DJ96's galaxies
(grey) are shown. The bulges appear to undergo slightly less evolution
in surface brightness than the disks and there is no apparent change in
the sizes of either component.
\label{magssizes}}
\end{figure}

We now turn to a key advantage of these AO images over previous galaxy
surveys -- the fact that all of the high-redshift bulges in our sample
are resolved. However, comparing these bulges to those nearby is not
straightforward.  Since the disks are the primary drivers of the
selection effects that define both distant and local samples, removing
these effects from the bulge statistics is difficult.  As a first
attempt, we chose actual galaxies out of DJ96's sample whose H-band disk
properties are consistent with galaxies in the selected, no-evolution
disk sample. We then compared the bulge properties of these galaxies to
the bulge properties of our sample.  In Figure~\ref{magssizes}, both
disk and bulge properties of this sub-sample of DJ's data and our
objects are plotted. The amount of bulge surface-brightness evolution
appears to be slightly less than that measured for the disks.  There are
some bulges whose surface brightness is higher than that of most local
galaxies, but this is a smaller fraction than for the disks. The sizes
of the bulges are within $\sim$5\% of local sizes and therefore also
show little evolution.  It is important to remember, however, that this
might be a biased comparison.  In order to quantify these conclusions
and put limits on the amount of bulge evolution consistent with our
data, we need to do a full, four-parameter analysis of all the selection
effects, which would require a larger sample size.

\section{Implications for Galaxy Evolution}

We have measured 0.6 mag arcsec$^{-2}$ of surface brightness evolution
through the use of IR, AO observations of galaxies at z$\sim$0.5. This
result is consistent with previous B-band results, mostly from HST
studies \citep[e.g.][]{schade, roche, lil, vogt, simard} although there
has been controversy over whether the evolution seen in these studies is
real or due to selection effects. We believe that we understand our
selection effects and that any influence they had on the samples were
removed.  By matching the selection effects of the two samples and
creating a Monte-Carlo, no-evolution sample we can directly measure what
the local population looks like when subjected to our selection
criterion, including removing any low-surface brightness galaxies we
would have missed. Even with this taken into account, there is a shift
in the average surface brightness of the disk population.  We therefore
believe that this evolution is real.

One caveat to this conclusion is that DJ96's sample included only
late-type spiral galaxies. Since our sample was not selected according
to Hubble type, we presumably include disk galaxies of all types. This
difference could lead to a bias in the comparison if the disk surface
brightness of the S0 and Sa galaxies, that we include but DJ96 don't, is
much higher than for the later types. Local samples, however, show that
this is not the case; there is very little change in the K-band, mean
disk surface brightness between early- and late-type spirals \citep{gra}.

Our results are also generally consistent with the models constructed by
\citet{bs}. They explore two types of models and compare
their predictions to the previous, B-band observations of disk surface
brightness evolution. The hierarchical models scale the galaxy disk
properties with the properties of the dark matter halos in which they
form.  The infall models concentrate on the conversion of gas into
stars, as constrained by observations of the Milky Way and other nearby
galaxies. The star formation rate is taken to fall off with galactic
radius and the luminosity of the galaxy at each radius is calculated
versus time from spectral synthesis codes.  Their models predict
$\sim$1.5 mag of B-band surface brightness evolution to z$\sim$1, which
they find is consistent with the level of evolution observed. However,
the models all give about the same predictions for the surface
brightness and size evolution, so it was not possible for them to
distinguish between the various models.

The fact that our measurement of disk surface-brightness evolution was
made in the IR which samples rest-frame optical light at high redshift,
rather than in the optical which samples rest-frame UV at high redshift,
significantly reduces the uncertainties in both the photometric and
morphological K corrections.  This data is less subject to the effects
of a recent burst of star formation and the measured shift more likely
represents a change in the underlying stellar population.  Although
\citet{bs} don't make any predictions of galaxy evolution when observed
in the IR, the level of evolution we see is generally consistent with
the previous data and with these models. \citet*{pozz} constructed a
simpler, pure luminosity evolution model from which they do make
predictions in the IR. Their K-band predictions, for an $\Omega$=0
universe, range from 0.4 mag of brightening to z=0.5 for a single-burst
population to 0.3 mag of fading to z=0.5 for a constant star-formation
model. The level of brightening we observe would clearly require some
amount of bursting star formation under this model.  Recent
stellar-population models, specifically designed for intermediate-age
populations and accounting in a self-consistent manner for AGB stars,
have suggested that these stars could lead to 0.3-0.5 mag of brightening
in the K band for populations 0.5-5 Gyr old, relative to models where
these stars are neglected \citep{mou1,mou2}.  This could reconcile
the \citet{pozz} models with our observations.

Even if the average properties of the high-redshift sample can't
constrain specific models of galaxy formation, it is clear that there is
a population of galaxies at high redshift that have higher surface
brightness than almost all galaxies seen locally.  From our Monte-Carlo
model, based on local galaxy properties, we would expect less than 20\%
of the disks in our sample to have a surface brightness above 16.5 mag
arcsec$^{-2}$, whereas we actually observe half of our objects above
this limit.  There is no reason why these bright disks wouldn't be
detected in the local sample since they have normal scale lengths and
high surface brightness.  This observation matches that made by both
\citet{vogt} and \citet{simard}, that there is a population of high
redshift galaxies with surface brightness higher than the locus of local
galaxies.  The most likely explanation for these objects is that they
have a younger stellar population than local galaxies and therefore
their surface brightness will fade with redshift and they will become
the normal galaxies seen today.  \citet{simard} suggest two other
possibilities to explain these objects: they are a type of galaxy at
high redshift that doesn't exist today or they are members of a rare
population that was therefore missed in the local sample (e.g.\
starburst HII galaxies). While we can't firmly rule these out until we
have better statistics, the passive evolution model is definitely the
simplest.

The lack of any observed, morphological evolution in the bulge
population is generally consistent with the simple picture that bulges
contain a uniformly old stellar population with little growth after an
initial burst. This is the first time the morphologies of bulges at high
redshift have been investigated, so there are no previous observations
or models to compare our results to. Moreover, since the bulge
evolutionary measurements we made were only qualitative, we can't make
any strong statements about their effect on the larger picture of bulge
evolution.

The data presented here are just the first part of a larger sample of
resolved, IR observations of distant galaxies. Future samples will
greatly increase the number of objects included and the quality of the
data. These observations should provide stronger constraints on the
amount of galaxy evolution taking place. Future models will hopefully be
more detailed and make specific predictions tailored to these
measurements. This combination should be a powerful tool in determining
how galaxies form and evolve.

\acknowledgments

We thank the Keck staff for creating the Keck AO facilities, without
which the observations reported here could not have been performed. We
thank Dr. Ian McLean and the NIRSPEC team for a great instrument and
early collaborations on this research. We also thank Dr. Robert Cousins
for useful discussions.

This work has been supported in part or full by the National Science
Foundation Science and Technology Center for Adaptive Optics, managed by
the University of California at Santa Cruz under cooperative agreement
No.  AST-9876783. Dr. James Larkin would also like to acknowledge the
Alfred P. Sloan foundation for their support through the faculty
fellowship program.

Data presented herein were obtained at the W.M. Keck Observatory, which
is operated as a scientific partnership among the California Institute
of Technology, the University of California, and the National Aeronautics
and Space Administration. The Observatory was made possible by the
generous financial support of the W.M. Keck Foundation.  The authors
wish to extend special thanks to those of Hawaiian ancestry on whose
sacred mountain we are privileged to be guests.  Without their generous
hospitality, none of the observations presented herein would have been
possible.

\clearpage

\epsscale{1.0}

\epsscale{1.0}

\begin{deluxetable}{lcccclcc}
\tabletypesize{\scriptsize}
\tablewidth{0pt}
\tablecolumns{6}
\tablecaption{AO Observations}
\tablehead{
\colhead{Object \tablenotemark{a}} & \colhead{K \tablenotemark{b}} & 
\colhead{Date \tablenotemark{c}}&\colhead{Band \tablenotemark{c}} & 
\colhead{Time (min) \tablenotemark{c}} & 
\colhead{\hspace{-1.5in}PSF star \tablenotemark{d}} &
\colhead{FWHM ($\arcsec$) \tablenotemark{e}} & \colhead{Strehl (\%) \tablenotemark{e}}
}
\startdata
PPM50296-7-23  & 16.7 & 1999 Dec & H & 35 & PPM127095-18+3 &0.13 &1.2 \\
PPM91714+22-19 & 17.0 & 1999 Sep & H & 35 & PPM91714+9-19 &0.16 &1.3 \\
PPM91714+22-19 & 17.0 & 2000 Aug & K$\arcmin$& 80 & PPM91714+9-19 &0.08 &5.6 \\
PPM102164+18-1 & 17.0 & 1999 Dec & H & 40 & PPM127095-18+3 &0.13 &1.2 \\
PPM102164+18-1 & 17.0 & 1999 Dec & J & 30 & PPM127095-18+3 &0.45 &0.2 \\      
PPM102164+18-1 & 17.0 & 2000 Jun & H & 48 & PPM102164-4+4/PPM106365-17+10 &0.04/0.06 &34.2/7.8 \\
PPM102164+18-1 & 17.0 & 2000 Jun & K$\arcmin$& 20 & PPM106365-17+10 &0.07 &8.1 \\
PPM102164+18-1 & 17.0 & 2000 Jun & J & 40 & PPM106365-17+10 &0.14 &0.9 \\        
PPM114182+6+27 & 17.1 & 1999 Sep & H & 35 & PPM114182+14+20 &0.05 &5.5 \\
PPM114182+6+27 & 17.1 & 2000 Jun & H & 48 & PPM114182+14+20 &0.05/0.06 &10.3/7.0 \\
PPM114182+6+27 & 17.1 & 2000 Jun & K & 20 & PPM114182+14+20 &0.06 &19.4 \\
PPM114182+6+27 & 17.1 & 2000 Jun & J & 20 & PPM114182+14+20 &0.09 &3.3 \\
PPM114182+6+27 & 17.1 & 2000 Aug & H & 55 & PPM114182+14+20 &0.14/0.11 &1.8/2.5 \\
PPM91088-21+18 & 17.6 & 1999 Sep & H & 45 & PPM91088-23+12 &0.07 &5.4 \\
PPM91088-21+18 & 17.6 & 1999 Dec & H & 15 & PPM91088-23+12 &0.19 &0.6 \\
PPM91714+18-17 & 17.6 & 2000 Aug & K$\arcmin$& 50 & PPM91714+9-19 &0.08 &5.6 \\
PPM106365-23+18& 17.7 & 2000 Jun & H & 20 & PPM106365-17+10 &0.06 &7.8 \\
PPM106365-23+18& 17.7 & 2000 Aug & K$\arcmin$& 50 & PPM106365-17+10 &0.07 &9.0\\       
PPM101029+20-8 & 18.1 & 2000 Jun & H & 30 & PPM101029+4-5/PPM106365-17+10 &0.04/0.06 &17.3/7.8 \\
PPM101029+20-8 & 18.1 & 2000 Jun & K$\arcmin$& 40 & PPM106365-17+10 &0.07 &8.1 \\
PPM102164+9+23 & 18.2 & 2000 Jun & H & 12 & PPM102164-4+4 &0.04 &34.2 \\
PPM127095-8+16 & 18.2 & 1999 Dec & H & 20 & PPM127095-18+3 &0.13 &1.2 \\
PPM127095-8+16 & 18.2 & 1999 Dec & J & 20 & PPM127095-18+3 &0.45 &0.2 \\
PPM127095-8+16 & 18.2 & 1999 Apr & H & 30 & PPM127095-18+3 &0.05 &5.4 \\
PPM115546-4+6  & 18.3 & 1999 Sep & H & 30 & PPM115546 &0.07 &6.5 \\
PPM115546-4+6  & 18.3 & 2000 Jun & H & 35 & PPM115546/PPM106365+7+6 &0.05/0.05 &17.3/18.5 \\
PPM115546-4+6  & 18.3 & 2000 Aug & K$\arcmin$& 40 & PPM115546 &0.06 &32.0 \\
PPM106365+4+13 & 18.4 & 2000 Jun & H & 60 & PPM106365-17+10 &0.06 &7.8 \\
PPM106365+4+13 & 18.4 & 2000 Aug & K$\arcmin$& 40 & PPM106365-17+10 &0.07 &9.0 \\
\enddata
\footnotesize
\tablenotetext{a}{The galaxies and PSF stars are named by the PPM
designation of their guide star plus the offset (in arcseconds) from
that star.} 
\tablenotetext{b}{K-band magnitude of the galaxy from the NIRC images.}
\tablenotetext{c}{The date, band, and exposure time (in minutes) of
the AO observations.}
\tablenotetext{d}{The PSF star used in analyzing each galaxy image.}
\tablenotetext{e}{The FWHM (in arcseconds) and Strehl ratio (in
  percentage) for each PSF star. Two values are given for those PSF
  stars for which there were two images at a given wavelength during the
  same observing run.}
\end{deluxetable}

\begin{deluxetable}{lllllc}
\tablewidth{0pt}
\tablecolumns{6}
\tablecaption{Derived Parameters}
\tablehead{
\colhead{Name} & \multicolumn{2}{c}{Disk} & \multicolumn{2}{c}{Bulge} & \colhead{Redshift}\\ 
\colhead{} & \colhead{$r_e$ \tablenotemark{a}} &  
\colhead{$\mu_o$ \tablenotemark{b}} & 
\colhead{$r_e$ \tablenotemark{a}} &
\colhead{$\mu_o$ \tablenotemark{b}} & \colhead{}
}
\startdata
PPM91714+22-19 &$0.81_{-0.19}^{+0.46}$ &$19.0_{-0.6}^{+0.8}$ &$0.119_{-0.022}^{+0.022}$ &$15.2_{-0.5}^{+0.5}$&0.46\\ 
PPM102164+18-1 &$0.29_{-0.07}^{+0.09}$ &$17.5_{-0.3}^{+0.4}$ &$0.053_{-0.017}^{+0.013}$ &$15.0_{-0.5}^{+0.5}$&0.55\\
PPM114182+6+27 &$0.43_{-0.08}^{+0.07}$ &$18.1_{-0.2}^{+0.1}$ &$0.045_{-0.009}^{+0.011}$ &$15.4_{-0.4}^{+0.5}$&0.48$\pm$0.05\tablenotemark{c}\\ 
PPM91088-21+18 &$0.49_{-0.13}^{+0.07}$ &$18.6_{-0.4}^{+0.1}$ &$0.084_{-0.015}^{+0.015}$ &$16.6_{-0.5}^{+0.5}$&0.70\\ 
PPM101029+20-8 &$0.24_{-0.08}^{+0.10}$ &$18.0_{-0.2}^{+0.3}$ &$0.025_{-0.009}^{+0.017}$ &$16.0_{-0.5}^{+0.7}$&0.6$\pm$0.3\tablenotemark{d}\\ 
PPM50296-7-23  &$0.38_{-0.15}^{+0.11}$ &$18.0_{-0.8}^{+0.2}$ &$0.028_{-0.015}^{+0.006}$ &$14.8_{-0.4}^{+0.5}$&0.6$\pm$0.3\tablenotemark{d}\\ 
PPM91714+18-17 &$0.55_{-0.32}^{+0.18}$ &$20.6_{-2.6}^{+0.2}$ &$0.130_{-0.054}^{+0.042}$ &$16.8_{-0.6}^{+0.7}$&0.46\\ 
PPM106365-23+18&$0.48_{-0.12}^{+0.62}$ &$19.2_{-0.4}^{+1.4}$ &$0.076_{-0.036}^{+0.144}$ &$18.2_{-0.8}^{+1.1}$&0.60\\ 
PPM127095-8+16 &$0.46_{-0.19}^{+0.59}$ &$19.2_{-0.8}^{+2.4}$ &$0.070_{-0.042}^{+0.200}$ &$17.6_{-0.9}^{+3.0}$&0.6$\pm$0.3\tablenotemark{d}\\ 
PPM106365+4+13 &$0.27_{-0.09}^{+0.49}$ &$19.5_{-0.4}^{+2.0}$ &$0.089_{-0.082}^{+0.090}$ &$20.0_{-1.8}^{+2.3}$&0.6$\pm$0.3\tablenotemark{d}\\ 
\enddata
\tablenotetext{a}{Exponential scale length of each component in arcsec.}
\tablenotetext{b}{Central surface brightness of each component in
magnitudes arcsec$^{-2}$.}
\tablenotetext{c}{Based only on a spectral break.}
\tablenotetext{d}{Based on the photometric limit of the sample.}
\end{deluxetable}


\newpage


\begin{thebibliography}{}

\bibitem[Bouwens \& Silk(2002)]{bs}{Bouwens, R. \& Silk, J. 2002, ApJ,
568, 522}

\bibitem[Bunker et al.(2000)]{bunk}{Bunker, A., Spinrad, H., Stern, D.,
Thompson, R., Moustakas, L., Davis, M., \& Dey, A. 2000, preprint
(astroph-0004348)}

\bibitem[Bureau(2002)]{bur}{Bureau, M. 2002, preprint (astroph-0203471)}

\bibitem[Cohen et al.(1999)]{coh99}{Cohen, J. G., Hogg, D. W., Pahre,
M. A., Blandford, R., Shopbell, P. L., \& Richberg, K. 1999, ApJS, 120,
171}

\bibitem[Cohen et al.(1996)]{coh96}{Cohen, J., Hogg, D. W., Pahre,
M. A., \& Blandford, R. 1996, ApJ, 462, L9}

\bibitem[Cohen(1994)]{mcoh}{Cohen, M. 1994, AJ, 107, 582}

\bibitem[Corbin et al.(2000)]{corb}{Corbin, M. R., O'Neil, E.,
Thompson, R. I., Rieke, M. J., \& Schneider, G. 2000, AJ, 120, 1209}

\bibitem[de Jong(1996)]{dj96}{de Jong, R. S. 1996, A\&A, 313, 45 (DJ96)}

\bibitem[Dinshaw et al.(1999)]{din}{Dinshaw, N., Evans, A. S., Epps,
  H., Scoville, N. Z., \& Rieke, M. 1999, ApJ, 525, 702}

\bibitem[Ellis, Abraham, \& Dickinson(2001)]{ell}{Ellis, R. S., Abraham,
  R. G., \& Dickinson, M. 2001, ApJ, 551, 111}

\bibitem[Graham(2001)]{gra}{Graham, A. W. 2001, MNRAS, 326, 543}

\bibitem[Larkin \& Glassman(1999)]{lg}{Larkin, J. E., \& Glassman,
T. M. 1999, PASP, 111, 1410}

\bibitem[Lilly et al.(1998)]{lil}{Lilly, S., et al. 1998, ApJ, 500, 75}

\bibitem[Mannucci et al.(2001)]{man}{Mannucci, F., Basile, F.,
Poggianti, B. M., Cimatti, A., Daddi, E., Pozzetti, L., \& Vanzi,
L. 2001, MNRAS, 326, 745}

\bibitem[Matthews \& Soifer(1994)]{ms}{Matthews, K., \& Soifer,
B. T. 1994, Exp. Astron., 3, 65}

\bibitem[McLean et al.(1998)]{mcl}{McLean, I. S., et al. 1998,
Proc. SPIE, 3354, 566}

\bibitem[Mouhcine \& Lan\c{c}on(2002a)]{mou1}{Mouhcine, M. \& Lan\c{c}on,
  A. 2002, A\&A, in press}

\bibitem[Mouhcine \& Lan\c{c}on(2002b)]{mou2}{Mouhcine, M. \& Lan\c{c}on,
  A. 2002, A\&A, in preparation}

\bibitem[Oke et al.(1995)]{oke}{Oke, J. B., et al. 1995, PASP, 107, 395}

\bibitem[Pozzetti et al.(1996)Pozzetti, Bruzual, \&
Zamorani]{pozz}{Pozzetti, L., Bruzual, G., \& Zamorani, G. 1996, MNRAS,
281, 953}

\bibitem[Press et al.(1996)]{press}{Press, W. H., Teukolsky, S. A.,
Vetterling, W. T., \& Flannery, B. P. 1996, Numerical Recipes in Fortran
77: The Art of Scientific Computing, Vol. 1 (2d ed.; Cambridge:
Cambridge University Press)}

\bibitem[Roche et al.(1998)]{roche}{Roche, N., Ratnatunga, K.,
Griffiths, R. E., Im, M., \& Naim, A. 1998, MNRAS, 293, 157}

\bibitem[Schade et al.(1996)]{schade}{Schade, D., Lilly, S., Le Fevre,
O., Hammer, F., \& Crampton, D. 1996, ApJ, 464, 79}

\bibitem[Simard et al.(1999)]{simard}{Simard, L., et al. 1999, ApJ,
519, 563}

\bibitem[Teplitz et al.(1998)]{tep}{Teplitz, H. I., Gardner, J. P.,
Malumuth, E. M., \& Heap, S. R. 1998, ApJ, 507, L17}

\bibitem[Vogt(1999)]{vogt}{Vogt, N. P. 1999, in ASP Conf. Ser. 193, The
  Hy Redshift Universe, ed. A. J. Bunker, \& W. J. M. van Breugel (San
  Francisco: ASP), 145}

\end{thebibliography}
\end{document}